\documentclass[tikz,nofootinbib,superscriptaddress,eqsecnum,tightenlines,11pt]{revtex4}

\usepackage{hyperref}
\usepackage{graphicx}
\usepackage{amsmath,amssymb,amsfonts,amsthm,stmaryrd,mathtools}
\usepackage{multirow,bigdelim}
\usepackage{dsfont}
\usepackage{color}
\usepackage{tikz}
\usepackage{tikz-3dplot}
\usetikzlibrary{calc}
\usetikzlibrary{shapes.geometric,patterns} 
\usepackage{blkarray}
\usepackage[normalem]{ulem}

\makeatletter
\newcounter{subsubsubsection}[subsubsection]
\renewcommand\thesubsubsubsection{\thesubsubsection .\@alph\c@subsubsubsection}
\newcommand\subsubsubsection{\@startsection{subsubsubsection}{4}{\z@}%
                                     {-3.25ex\@plus -1ex \@minus -.2ex}%
                                     {1.5ex \@plus .2ex}%
                                     {\centering\normalfont\small\textit}}
\newcommand*\l@subsubsubsection{\@dottedtocline{3}{10.0em}{4.1em}}
\newcommand*{\subsubsubsectionmark}[1]{}
\makeatother
\newcommand\scalemath[2]{\scalebox{#1}{\mbox{\ensuremath{\displaystyle #2}}}}

\def\be{\begin{equation}}
\def\ee{\end{equation}}
\def\ba{\begin{eqnarray}}
\def\ea{\end{eqnarray}}

\newcommand\nn{\nonumber}
\newcommand{\q}{\quad}
\newcommand{\Lap}{\scalemath{0.75}\Delta_{\rm b}}

\newcommand{\im}{\mathrm{i}}

\newcommand{\p}{{\partial}}

\def\b{{}^{\rm b}\!}

\newcommand{\qq}{{\qquad}}

\begin{document}

\title{  Perfect discretizations as a gateway to  one-loop partition functions for  4D gravity  }

\author{Seth K. Asante}
\email{sasanteATuni-jena.de}
\affiliation{Perimeter Institute for Theoretical Physics, 31 Caroline Street North, Waterloo, ON, N2L 2Y5, Canada}
\affiliation{Theoretisch-Physikalisches Institut,
Friedrich-Schiller-Universit\"at Jena,
Max-Wien-Platz 1,
07743 Jena, Germany}

\author{Bianca Dittrich}
\email{bdittrichATperimeterinstitute.ca}
\affiliation{Perimeter Institute for Theoretical Physics, 31 Caroline Street North, Waterloo, ON, N2L 2Y5, Canada}
\affiliation{Institute for Mathematics, Astrophysics and Particle Physics,
Radboud University, Heyendaalseweg 135, 6525 AJ Nijmegen, The Netherlands}

\begin{abstract}

%Renormalization procedure for 4D cylinder space--time
%We shall discuss the interplay between holography  duality and boundary degrees of freedom also known as boundary gravitons in 3+1 dimensions for finite boundaries. We shall  describe a boundary theory for these boundary gravitons which can be encoded in geodesics lengths normal to the boundary (at least for flat space--times) space--time. We shall also use renormalization techniques to compute the Hamilton Jacobi functionals and one loop partition functions for gravity. These renormalization techniques connect computations performed in both discrete and continuum space--times. 

Lattice actions and amplitudes that perfectly mirror continuum physics are known as perfect discretizations. Such perfect discretizations naturally preserve the symmetries of the continuum. This is a key concern for general relativity, where diffeomorphism symmetry and dynamics are deeply connected, and diffeomorphisms play a crucial role in quantization. In this work we construct for the first time a perfect discretizations for four-dimensional linearized gravity. We show how the perfect discretizations do lead to a straightforward construction of the one-loop quantum corrections for manifolds with boundary. This will also illustrate, that for manifolds with boundaries, gauge modes that affect the boundary, need to be taken into account for the computation of the one-loop correction. This work provides therefore an evaluation of the boundary action for the diffeomorphism modes for a general class of backgrounds.
 
\end{abstract}

\maketitle

%\newpage

\section{Introduction}

Discretizations are a well utilized method to regularize  and define path integrals and thus the quantization of a given system. Applied to general relativity discretizations generically break, however, the fundamental symmetry of the theory --- diffeomorphism symmetry \cite{BahrDittrich09a}. This has been a key obstacle for the construction of a consistent theory of quantum gravity \cite{DittrichReview14,ThiemannRenRev}.

The reason for the breaking of diffeomorphism symmetry is that the action of a diffeomorphism in the discrete can affect the discretization itself \cite{BahrDittrich09a,DittrichReview12}. E.g. one way to discretize a metric manifold is via an embedded lattice, whose edges approximate geodesics between the lattice's neighbouring vertices. The gravitational variables determine the lengths of these edges and characterize in this way the discretization. 
             
Diffeomorphisms act on a smooth manifold by moving points. In the kind of discretization described above diffeomorphisms move the vertices of the grid and change in this way the geodesics and their lengths and thus the discretization itself.  Vertices can be even moved on top of each other and in this manner even coarsen the lattice \cite{BDS11}.

A system, for which diffeomorphism symmetry holds, would give the same predictions for different lattices related by this action (for observables that are supported by these different lattices). As this action can relate coarser and finer lattices, predictions derived from any given lattice have to coincide with predictions from any refinement of this lattice, including  from the infinite refinement or continuum limit. 

A family of discretizations, whose members encode the same predictions as the continuum (limit), is called perfect \cite{Hasenfratz,Improve}. Perfect actions have been proposed in Lattice QCD as one method to restore Lorentz symmetry in the discrete \cite{Hasenfratz}. In the case of gravity or reparametrization invariant systems, perfect actions do restore diffeomorphism \cite{Improve} or reparametrization invariance \cite{BahrDittrich09b,BDS11}, respectively.

By definition, a perfect action for a coarser lattice can be re-constructed from the perfect action of a finer lattice by integrating out the finer degrees of freedom. Thus perfect discretizations can be constructed by blocking from the continuum \cite{Bietenholz,BDH,Liegener} or alternatively,  via an iterative procedure, in which the lattice is repeatedly refined and then coarse grained by integrating out the finer degrees of freedom \cite{Improve,BDS11}. 

This iterative procedure does define a renormalization flow for the amplitude, or on the classical level, the action. Its fixed point is a candidate for a perfect action or discretization.  The fixed point conditions itself can be also used directly to determine a perfect discretization, which determines the solution to the continuum path integral itself. This has been illustrated for quantum mechanical path integrals in \cite{BDS11}. 

We will generalize this strategy to a quantum field theory system, and show how the one-loop correction can be determined from such a fixed point condition. This construction, although limited to a certain class of backgrounds\footnote{The background manifold with boundary ${\cal M}$ needs to be such, that gluing two such manifolds we obtain a manifold with the same type of background. I.e. gluing a spherical shell with outer and inner radius $(r_3,r_2)$ to another spherical sphere with $(r_2,r_2)$ gives again a spherical sphere with radii $(r_3,r_2)$.}, allows a  straightforward determination of the one-loop correction for manifolds with boundaries. 

This computation of the one-loop correction requires the determination of the perfect action for linearized gravity. In this work we therefore determine --- to our knowledge for the first time --- the perfect action for a dynamical system with propagating (bulk) gravitons, here linearized gravity in four spacetime dimensions. Previous work \cite{BDS11} considered only scalar fields, vector fields and gravity in three dimensions (which does not feature bulk gravitons),  \cite{Liegener} does also discuss  scalar and vector fields, \cite{Bietenholz} considers in addition fermionic systems.

To utilize the power of the fixed point equations and at the same time the fact that we are considering a free system, we employ a real space discretization in only one direction, which here we will choose as radial direction. For the other three directions we employ a Fourier transform. The Fourier modes do decouple in their dynamics (in radial direction). We can thus easily impose a cut-off, and see this as a further discretization \cite{Kempf,DS13}, or take all modes into account and work with a system that is continuous in three of its dimensions and discrete in the remaining one.
The Fourier transform does allow us to deal with the non-locality of the one-loop correction, that has to be expected to appear for systems with propagating degrees of freedom, including four-dimensional gravity \cite{Meas1}.

A further technique we are employing is the split between diffeomorphism or gauge modes and (bulk) graviton modes. The diffeomorphism modes can be localized\footnote{This holds with the exception for the lowest Fourier modes, see \cite{SBF} for a detailed discussion.} to a given boundary component, whereas the bulk gravitons do lead to non-trivial couplings. Importantly, the contribution from the diffeomorphism modes cannot be neglected: they do contribute non-trivially to the Hamilton-Jacobi function for linearized gravity in case that we do have boundaries. The Hamilton-Jacobi functional provides  the (classical part of the) perfect action. In fact \cite{Hol1,SBF} consider three-dimensional gravity for manifolds with boundary, which does not feature bulk gravitons, but nevertheless result in non-trivial Hamilton-Jacobi functionals and one-loop corrections. The latter have been previously computed in \cite{OneLoop} using heat kernel methods. The works \cite{Hol1,SBF} show that  the diffeomorphisms normal to the boundary (also referred to as boundary gravitons) can be alternatively described as a scalar field on the boundary, with a Liouville like coupling to the boundary Ricci-scalar. These systems are therefore examples for a holographic duality, which holds for finite boundaries and can be also extended to the non-perturbative level \cite{HolPR}.

Using discrete (Regge) gravity the work \cite{Hol4D} showed that this dual description of boundary gravitons generalizes also to certain four-dimensional backgrounds and leads to a similar one-loop correction as for the corresponding three-dimensional example. The work \cite{Hol4D} restricted however to a discretization without refinement in radial direction. This allowed to capture the diffeomorphism modes, but not the bulk gravitons. The current work does provide a completion of \cite{Hol4D}, as it also includes the bulk graviton contributions to the partition function.

The contribution of the diffeomorphism modes relies on a generalization of results from \cite{SBF} from three dimensions to general dimensions, which we  present in Appendix \ref{AppHJF}. This generalization does also clarify for which four-dimensional backgrounds we obtain a dual description in terms of a {\it local} field theory on the boundary. 

~\\
This paper is organized as follows: In Section \ref{SecFlat} we will introduce the background geometry, four-dimensional spinning thermal flat space, which we will consider here, as well as further notions like the twisted Fourier transform, needed in the remainder in the work. In Section \ref{SecHJF} we will compute the Hamilton-Jacobi functional, that is the action evaluated on solutions. Here we will consider separately the contributions from the graviton modes and the diffeomorphism modes. For the latter we provide an evaluation applicable to a large family of backgrounds of general dimensions, in Appendix \ref{AppHJF}. In Section \ref{HJRec} we explain how the graviton contribution to the Hamilton-Jacobi functional can be alternatively determined using recursion relations or fixed point equations.  The fixed point equations are then used to construct the one-loop correction in Section \ref{SecOLC}. We close with a discussion and outlook in Section \ref{Disc}.

\section{Twisted thermal  flat space with finite boundary}\label{SecFlat}

The background geometry, which we will consider here, is a generalization of the twisted or spinning thermal flat space \cite{TwistedFlatSpace} from three to four dimensions, and a solution to Einstein's equation (with Euclidean signature) without cosmological constant. The `twisting' introduces moduli parameters, that turn out to lead to an interesting structure for the one-loop correction.

 The metric $g_{ab}$ of thermal spinning flat space in four dimensions is given by
\ba\label{metric0TSF}
 g_{ab} dx^a dx^b=  dr^2+ r^2 d\theta^2 + dt^2 + dz^2
\ea
with the coordinates subject to periodic identifications
\ba \label{periods}
(r,\theta,t,z)&\sim& (r,\theta+\gamma_t, t + \beta_t, z) \nn \\
(r,\theta,t,z)&\sim& (r,\theta+\gamma_z, t , z+\beta_z)\q ,
\ea
in addition to the usual identification $(r,\theta,t,z)\sim(r,\theta+2\pi,t,z)$ for the angular variable $\theta$ and the condition $r\geq 0$ for the radial coordinate. The space time is flat, but features moduli parameters $(\beta_t,\gamma_t,\beta_z,\gamma_z)$.
 The background extrinsic curvature for the $r=\text{const}$ hypersurfaces has only one non-vanishing component $K_{\theta \theta}=r$.

We will consider perturbations of this background metric
\be\label{MetricP}
g_{ab}^{\rm full} = g_{ab} + \gamma_{ab} \q .
\ee
Background quantities will be denoted with latin letters (except for the Christoffel symbols which will be denoted by $\Gamma$), perturbative quantities will be denoted by small case greek letters or calligraphic script letters.

We will consider two different choices of boundaries. For the first choice we just have an `outer' boundary at $r=r_{\rm out}$, thus we have  $r\in [0,r_{\rm out}]$ for the radial coordinate.  This gives a manifold with the topology of a solid 3-torus.  The second choice is with an outer boundary at $r=r_{\rm out}$ and an inner boundary at $r=r_{\rm in}<{\rm r_{\rm out}}$, hence $r\in [r_{\rm in},r_{\rm out}]$, and we will refer to the resulting topology as toroidal annulus. A property of these annuli is that one can glue two of these together (if the outer radius of one matches the inner radius of the other), and obtain again a toroidal annulus. This gluing operation can be used to define recursion relations, that can be used to determine the path integral for the toroidal annulus, along the lines of \cite{BDS11}.

\begin{figure}[ht!]
\begin{center}
\begin{tikzpicture}
\begin{scope}[xshift=-6cm,line width=1.pt,color=black,scale = 2.4]
  \draw [fill=gray!50]
  (180:5mm) coordinate (a)
  -- ++(0,-12.5mm) coordinate (b) node [midway, left, inner sep=2pt] {$\beta$}
  arc (180:360:5mm and 1.75mm) coordinate (d)
  -- (a -| d) coordinate (c) arc (0:180:5mm and 1.75mm);
  \draw [fill=gray!70]
  (0,0) coordinate (t) circle (5mm and 1.75mm);
  \draw [thick,->] (0.55,0) to [bend right] node [midway,  right] {$\gamma$} (0.4,0.15) ;
  \draw [densely dashed,black!60,line width=0.2pt] (d) arc (0:180:5mm and 1.75mm);
  \fill[red] (-0.32,-0.13)circle (0.6pt) ;
  \fill[red] (0.01,-1.43)circle (0.6pt) ;
\end{scope}
\begin{scope}[scale = 2.4]
 \draw [fill=gray!50,thick]
  (180:5mm) coordinate (a) -- ++(0,-12.5mm) coordinate (b) node [midway, left, inner sep=2pt] {$\beta$}
  arc (180:360:5mm and 1.75mm) coordinate (d)-- (a -| d) coordinate (c) arc (0:180:5mm and 1.75mm);
  \draw [fill=gray!70,thick]  (0,0) coordinate (t) circle (5mm and 1.75mm);
  \draw [thick,->] (0.55,0) to [bend right] node [midway,  right] {$\gamma$} (0.4,0.15) ;
  \draw [densely dashed,black!50,line width=0.2pt] (d) arc (0:180:5mm and 1.75mm);
   \draw [fill=white!30]  (t) circle (2mm and 0.7mm);
   \draw[black!70]  (0,-1.25) circle (2mm and 0.7mm);
   \draw[black!70] (2mm,-1.25)-- (2mm,-0.16) (-2mm,-1.25)-- (-2mm,-0.16) ;
    \fill[red] (-0.32,-0.13)circle (0.6pt) ;
  \fill[red] (0.01,-1.43)circle (0.6pt) ;
 \end{scope}
 \begin{scope}[xshift = 6cm,scale = 2.4]
 \draw [fill=gray!50,thick]
  (180:5mm) coordinate (a) -- ++(0,-12.5mm) coordinate (b) node [midway, left, inner sep=2pt] {$\beta$}
  arc (180:360:5mm and 1.75mm) coordinate (d)-- (a -| d) coordinate (c) arc (0:180:5mm and 1.75mm);
  \draw [fill=gray!70,thick]  (0,0) coordinate (t) circle (5mm and 1.75mm);
  \draw [fill=gray!50]  (0,0) coordinate (t) circle (3.4mm and 1.2mm);
  \draw [thick,->] (0.55,0) to [bend right] node [midway,  right] {$\gamma$} (0.4,0.15) ;
  \draw [densely dashed,black!70,line width=0.2pt] (d) arc (0:180:5mm and 1.75mm);
   \draw [fill=white!30]  (t) circle (2mm and 0.7mm);
   \draw[black!70]  (0,-1.25) circle (2mm and 0.7mm);
   \draw[black!70]  (0,-1.25) circle (3.4mm and 1.2mm);
   \draw[black!70] (2mm,-1.25)-- (2mm,-0.16) (-2mm,-1.25)-- (-2mm,-0.16) ;
   \draw[black!70] (3.4mm,-1.25)-- (3.4mm,-0.15) (-3.4mm,-1.25)-- (-3.4mm,-0.15) ;
    \fill[red] (-0.32,-0.13)circle (0.6pt) ;
  \fill[red] (0.01,-1.43)circle (0.6pt) ;
    \fill[red] (-0.32,-0.13)circle (0.6pt) ;
  \fill[red] (0.01,-1.43)circle (0.6pt) ;
 \end{scope}
\end{tikzpicture}
\caption{[From left to right] A solid cylinder, a cylindrical annulus and two annuli glued together in three dimensions. After rotating the tops by a twist angle, they are identified with the bottoms (red dots), leading to a solid torus and toroidal annuli.} \label{4flattorus}
\end{center}
\end{figure}
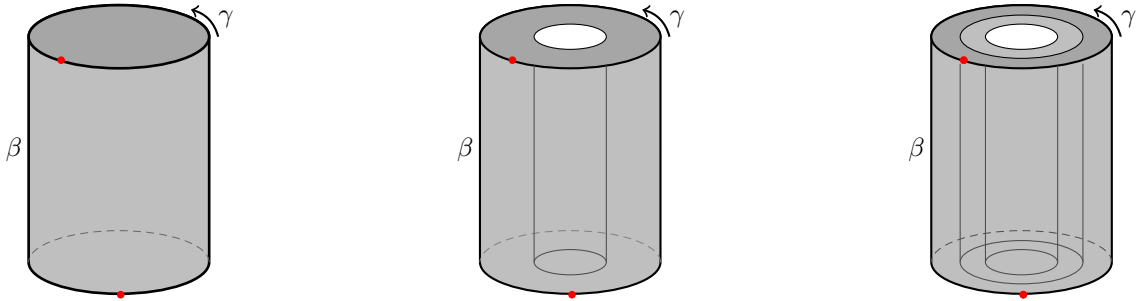

The slicing in the radial coordinate $r$ will therefore play an important role here. We use coordinates $x^a=(r,y^A)$ for this slicing, with capital latin indices $A$ indicating the coordinates $y^A=(\theta,t,z)$.

%If we consider the space--time for $0\leq r\leq r_{\rm out}$ we obtain a solid 3--torus. Contractible cycles include curves described by $t=\text{const},\, r=\text{const}$ and $z=\text{const},\, r=\text{const}$. Non--contractible cycles include curves along  $\theta=\text{const}., \, r=\text{const}$. The 3--torus can be obtained by identifying the boundary of the hyper--cylinder of height $\beta_t$ with a twisting angle (or angular potential) $\gamma_t$ and afterwards identify the boundary resulting torus with height $\beta_z$, with a twisting angle  $\gamma_z$.   

The boundary at $r=\text{const.}$ has the topology of a 3--torus and is flat (with respect to the background metric) ${}^3\!R_{ABCD}=0$.  This allows us to define an (almost) standard Fourier transform for the metric perturbation components. We need however to adjust the Fourier transform to our periodic identifications   \eqref{periods}. This leads to a  `twisting'  of the phase factors \cite{Hol1}, so that these are well defined on the manifold specified by the periodic identifications in \eqref{periods}. The Fourier-transformed metric perturbations are then given by 
 \ba\label{Ftrafo1}
 \gamma_{ab}(r,k_\theta,k_t,k_z)&=& \frac{1}{\sqrt{2\pi \beta_t \beta_z}} \int_{0}^{\beta_t}   dt \, \int_{0}^{\beta_z} dz  \int^{2\pi}_{0}   d\theta \, \,\gamma_{ab}(r,\theta,t,z) \,e^{ - \im \left( \theta k_\theta+ t  k_t+z k_z \right)} \q ,
 \ea
 where we have introduced the abbreviations
 \be \label{35}
 k_t:=\frac{2\pi }{\beta_t} ( k'_t-\frac{\gamma_t}{2\pi} k_\theta), \qq  k_z:=\frac{2\pi }{\beta_z} ( k'_z-\frac{\gamma_z}{2\pi} k_\theta)
 \ee 
 with $k_\theta,k'_t,k'_z \in {\mathbb Z}$. The inverse transform is given by
 \ba
 \gamma_{ab}(r,\theta,t,z)&=& \frac{1}{\sqrt{2\pi \beta_t\beta_z}}  \sum_{k_\theta,k'_t,k'_z} \gamma_{ab}(r,k_\theta,k_t,k_z)  \,\,e^{ \im \left( \theta k_\theta+ t  k_t+z k_z \right)} \q .
 \ea

We thus have for the inner product between the metric perturbations
 \ba\label{innerprod}
 \langle \gamma , \gamma' \rangle := \frac{1}{{}^3\!V} \int d^3y \, \sqrt h \,  \gamma_{AB} h^{AC} h^{BD} \gamma'_{CD}  
 \,=\, \frac{1}{2\pi \beta_t\beta_z} \sum_{k_\theta,k'_t,k'_z}  \,  \gamma_{AB}(k) h^{AC} h^{BD} \gamma'_{CD}(-k) 
 \ea
 where ${}^3\!V:= \int d^3y \, \sqrt h  =2\pi \beta_t\beta_z r$ is the volume for the hypersurface.

\section{ The Hamilton-Jacobi functional}\label{SecHJF}

As part of the one--loop path integral we need to determine the Hamilton-Jacobi functional, that is the action evaluated on solutions. We consider Dirichlet boundary conditions, that is the induced metric perturbations $\gamma_{AB}$ on the boundary are kept fixed. The Hamilton-Jacobi functional does thus depend on the induced metric perturbations $\gamma_{AB}$ at $r=r_{\rm out}$, if we only have an outer boundary, or, at $r=r_{\rm out}$ and at $r=r_{\rm in}$, if we have an outer and an inner boundary. % In the first case one needs to assume some smoothness conditions at $r=0$, which {\color{blue} will be discussed later}.

We will compute the Hamilton-Jacobi functional to second order in the metric perturbations $\gamma_{AB}$. The second order Hamilton-Jacobi functional is also needed to determine the one-loop determinant (for the toroidal annulus topology) via a recursion relation.

The zeroth and first order of the Hamilton-Jacobi functional result from the Gibbons-Hawking-York boundary term (for the case without cosmological constant) and can be determined without solving equations of motions for the metric perturbations. 

The second order of the Hamilton-Jacobi functional splits into three types of contributions, which result from a splitting of the metric perturbations into (linearized) diffeomorphisms and gravitons.  We will describe this split in more detail in section \ref{split}. The three types of terms are: $(i)$ one term is quadratic in perturbations describing (linearized) diffeomorphisms of the background space time, which affect the boundary, $(ii)$ another term is quadratic in perturbations describing (bulk) gravitons, and $(iii)$ a third couples the diffeomorphism sector and the graviton sector.  Only the term quadratic in the gravitons involves terms that couple fields from both boundaries (in the two-boundary case). It is also the only term, whose computation does involve solving a differential equation in the radius.

In particular the term quadratic in perturbations describing diffeomorphisms can be computed for general spacetimes with boundaries. This determines the (second order of the) Hamilton-Jacobi functional for flat solutions, but with boundaries, that are allowed to `fluctuate'.  In Appendix \ref{AppHJF} we will present a systematic evaluation of this part of the second order functional.

\subsection{Zeroth and first order}

Here we will compute the background value and the first order of the Hamilton-Jacobi functional (HJF). We will denote the HJF by $S_{\rm HJoi}$ for the case with an outer and inner boundary, and by $S_{\rm HJo}$ if we have only one boundary. The expansion  is denoted by $S_{\rm HJ}=S_{\rm HJ}^{(0)}+S_{\rm HJ}^{(1)}+\cdots$.

The (Euclidean) Einstein--Hilbert action, with Gibbons-Hawking-York boundary term, is given by
\ba\label{actionC1}
S&=& -\frac{1}{2\kappa} \int_{\cal M} d^{d} x \, \sqrt{g}  \left( 
R 
 - 2 \Lambda \right)
 \, \,\,-\,\,\frac{1}{\kappa} \int_{\partial \cal M} d^{(d-1)} y \sqrt{h}  \, \epsilon K   \q ,
\ea 
where $\kappa=8\pi G_N$ and $G_N$ is Newton's constant.  The manifold ${\cal M}$ is described in section \ref{SecFlat} and the boundary $\delta {\cal M}$ is either given by one component (at constant radius $r_{\rm out}$) or two components (at constant radii $r_{\rm out}$ and $r_{\rm in}$), with $h_{AB}$ denoting the intrinsic metric of the boundary.
We will also use the following convention regarding the (sign of the) extrinsic curvature tensor: $K_{AB}$ will be understood as the extrinsic curvature tensor associated to the foliation of ${\cal M}$ by surfaces of constant radius, that is, defined by $K_{AB}=\tfrac{1}{2}\partial_r h_{AB}$, where $h_{AB}$ is the induced metric on a $r=\text{const.}$ hypersurface. This  agrees with the orientation of the outer boundary, for which we define $\epsilon=+1$. For the inner boundary, we have however an  outward pointing normal $n^a=(-1,0,\ldots)$, and therefore $\epsilon=-1$.

Without a cosmological constant, only the boundary term contributes to the Hamilton-Jacobi functional. For the background metric (\ref{metric0TSF}) and an $r=\text{const.}$ hypersurface the extrinsic curvature is given by $K_{AB}= r\delta^\theta_A\delta^\theta_B$ and the trace is given by $K=1/r$. With $\sqrt{h}=r$ we see that the contribution from one boundary component does not depend on the value of  $r_{\rm out}$ or $r_{\rm in}$. 

For the case with an outer ans inner boundary the two contributions cancel, and the zeroth order of the HJF vanishes $ S^{(0)}_{\rm HJoi}=0$. If we have only an outer boundary, the zeroth order is given by
\ba
S^{(0)}_{\rm HJo}&=&-\frac{2\pi\beta_t\beta_z}{\kappa} \q,
\ea 
which thus depends only on the moduli parameters $\beta_t,\beta_z$, but not on $\gamma_t, \gamma_z$.

The first variation of the action  (\ref{actionC1}) is given by
\ba\label{C17}
-\kappa  S^{(1)} &=&
\tfrac{1}{2}\!\!\int_{\cal M} \! d^d x \, \sqrt{g}\left( \left( \tfrac{1}{2}R -\Lambda \right) g^{ab} -R^{ab}\right) \gamma_{ab} +
%\nn\\&&
\tfrac{1}{2} \!\! \int_{\partial \cal M}\! \!\! \! d^{(d-1)} \! y \, \sqrt{h} \epsilon \left(   K h^{AB}\  -  K^{AB}   \right)\gamma_{AB}\, , \q\q
\ea
where the bulk term determines the (background) equations of motion and the boundary term gives the first order of the HJF. The boundary term does only involve the induced metric perturbations at the boundary and no (radial) derivatives thereof. It can therefore be evaluated without solving any equations of motion and gives for the background metric (\ref{metric0TSF})  
\ba
-\kappa S^{(1)}_{\rm HJoi}&=& \tfrac{1}{2} r_{\rm out} \int d\theta dtdz \,(\gamma_{tt}+\gamma_{zz})(r_{\rm out}) \,\,-\,\,  \tfrac{1}{2} r_{\rm in} \int d\theta dtdz \, (\gamma_{tt}+\gamma_{zz})(r_{\rm in}) \nn\\
%&=& \tfrac{1}{2} \sqrt{2\pi \beta_t \beta_z} \left( r_{\rm out} (\tilde \gamma_{tt}+\tilde \gamma_{zz})(r_{\rm out})- r_{\rm in} (\tilde \gamma_{tt}+\tilde \gamma_{zz})(r_{\rm in})\right) \nn\\
-\kappa S^{(1)}_{\rm HJo}&=& \tfrac{1}{2} r_{\rm out} \int d\theta dtdz \,(\gamma_{tt}+\gamma_{zz})(r_{\rm out}) \q .
%
%=\tfrac{1}{2} \sqrt{2\pi \beta_t \beta_z} \, r_{\rm out} \,(\tilde \gamma_{tt}+\tilde \gamma_{zz})(r_{\rm out})
\ea
This first order contribution does vanish, if the metric perturbation arises purely from a (linearized) diffeomorphisms -- in this case the integrands due amount to total divergencies.  This is generally the case for metric perturbations describing a diffeomorphism whose generating vector field at the boundary is tangential to this boundary. We will refer to such diffeomorphisms as boundary tangential diffeomorphisms, and the vanishing of the first order HJF for such metric perturbations reflects the invariance of the full HJF under boundary diffeomorphisms. 

For the background space time and boundary considered here, the first order contribution also vanishes for diffeomorphisms, whose generating vector field is normal to the boundary, and which can therefore interpreted to move the boundary.  These boundary normal diffeomorphisms are sometimes referred to as boundary gravitons (not to be confused with bulk gravitons). 

Note that metric perturbations describing tangential diffeomorphism will contribute to the second order of the HJF. This does not contradict the invariance of the (full) HJF under tangential diffeomorphisms -- rather, these second order terms are needed to render the first order terms invariant to higher order.

\subsection{Second order: diffeomorphism and graviton modes}\label{split}

The second order HJF encodes the linearized dynamics. But there are different types of contributions to the HJF, which arise from the split of the metric perturbations into those resulting from linearized diffeomorphisms and those describing (bulk) gravitons. 

Solutions -- modulo diffeomorphisms -- are determined by the boundary data, that is the   boundary  metric perturbations. The split into diffeomorphism and graviton sectors can therefore be formulated via conditions on the  boundary metric perturbations. We can describe the sectors via projectors, which are orthogonal with respect to the inner product (\ref{35}) for the boundary metric perturbations.  

In the following we will exclude\footnote{The summation $\sum_{k_\theta,k'_t,k'_z}$ does also not include $(k_\theta,k_t',k_z')\neq (0,0,0)=0$.} the case $k_\theta=k'_t=k'_t=0$, as it requires a separate treatment, see appendix \ref{zeromode}. We also assume that $k_t^2+k_z^2\neq 0$. (We remind the reader, that $k_t=\tfrac{2\pi}{\beta_t}(k_t'-\tfrac{\gamma_t}{2\pi} k_\theta)$ and $k_z=\tfrac{2\pi}{\beta_z}(k_z'-\tfrac{\gamma_z}{2\pi} k_\theta)$ with $(k_\theta,k_t',k_z')\in \mathbb{Z}^3$. ). This is assured for all $(k_\theta,k_t',k_z')\neq (0,0,0)$ if either $\gamma_t/2\pi$ or $\gamma_z/2\pi$ is irrational.

The sector describing (pure) diffeomorphisms is spanned by boundary perturbations of the form
\ba\label{Dif1}
 \zeta_{AB}&=& [{\cal L}_\xi g]_{AB}\,=\,2 K_{AB} \xi_r + D_A \xi_B + D_B \xi_A  \q 
 \ea
 For our case, and in the Fourier transformed picture we obtain
 \ba\label{Dif1F}
  \zeta_{AB} &=& 2 r \delta^\theta_A\delta^\theta_B  \xi^r + \im k_A \xi_B+  \im k_B \xi_A  \q ,
\ea
with $\xi^r$ parametrizing boundary normal diffeomorphisms and $(\xi^\theta,\xi^t,\xi^z)$ parameterizing boundary tangential diffeomorphisms.

The graviton sector is spanned by boundary metric perturbations of the form 
\ba\label{gravi1}
 \chi_{AB} =  w  W_{AB} +  x  X_{AB}
\ea
where  
 \ba\label{gravi1a}
(W_{\theta\theta}, W_{tt}, W_{zz}, W_{\theta t}, W_{\theta z}, W_{tz}) &=& \frac{1}{(k_t^2+k_z^2)} \left( 0, k_z^2, k_t^2, 0,0, -k_t k_z \right) \nn \\
(X_{\theta\theta}, X_{tt}, X_{zz},X_{\theta t}, X_{\theta z}, X_{tz}) &=& \frac{1}{r}\sqrt{\frac{2}{\Lap}} \left( 0, -\frac{k_\theta k_t k_z}{ (k_t^2+k_z^2)}, \frac{k_\theta k_t k_z}{ (k_t^2+k_z^2)}, \frac{r^2 k_z}{2}, -\frac{r^2k_t}{2} , \frac{k_\theta( k_t^2 - k_z^2)}{2(k_t^2+k_z^2)} \right) \nn\\
\ea
and $\Lap := \frac{k_\theta^2}{r^2}+k_t^2+k_z^2$. The perturbations $W_{AB}$ and $X_{AB}$ are orthogonal to each other, as well as orthogonal to the diffeomorphism sector (\ref{Dif1}), with respect to the inner product (\ref{innerprod}).   

Thus, the $x$ and $w$ parameters are given by 
 \ba
w[\gamma] :=\frac{ \langle \gamma, W \rangle }{\langle W, W\rangle} &=& \frac{1}{(k_t^2+k_z^2)} \left( k_z^2 \gamma_{tt} + k_t^2 \gamma_{zz} - 2k_tk_z \gamma_{tz} \right) \q, \nn \\
x[\gamma] := \frac{\langle \gamma, X \rangle }{\langle X, X\rangle }&=& \frac{1}{r}\sqrt{\frac{2}{\Lap}} \left( \frac{k_\theta k_t k_z}{k_t^2+k_z^2} (\gamma_{zz}-\gamma_{tt}) +k_z \gamma_{\theta t} -k_t \gamma_{\theta z} + \frac{k_\theta (k_t^2-k_z^2)}{k_t^2+k_z^2} \gamma_{tz} \right) \q.
\ea

We have inverse differential operators (acting along a boundary component) appearing in $W_{AB}$ and $X_{AB}$, the projectors constitute therefore non-local operators. The projectors can be restricted to either the outer or inner boundary component. However, to suppress bulk gravitons, we need to demand that $w=0$ and $x=0$ for both boundaries. Of course, bulk diffeomorphisms can appear even if we set the $\xi$--parameters at both boundaries to zero. The HJF will however only depend on the boundary diffeomorphisms, and their contribution factorizes over the boundary components.

\subsection{Equations of motion for the metric perturbations}

The second order expansion of the action (\ref{actionC1}) can be obtained with a somewhat lengthy calculation, see \cite{SBF}. It is of the form\footnote{The split into bulk and boundary terms is not uniquely determined, as one can redefine this split using integration by parts. Here we have chosen a form, where the bulk term vanishes on-shell.} 
\ba\label{S2O}
-\kappa S^{(2)}&=& \frac12 \int d^d x \sqrt{g} \gamma_{ab} \hat G^{ab} \,\,+\,\, \frac12 \int d^{(d-1)} \sqrt{h}\,  \epsilon \,\,\gamma_{ab} \left( (B_1)^{abcd} \gamma_{cd} + (B_2)^{abcde} \nabla_c \gamma_{de}\right)
\ea 
where $\hat G^{ab}$ is linear in the metric perturbations $\gamma_{ef}$ and, together with $B_1$ and $B_2$, is detailed in appendix \ref{App2ndorder}.

The variation of the bulk term with respect to the metric perturbations yields the equation of motions 
\ba\label{eom01}
\hat G^{ab}=0
\ea
for the linearized theory.  Due to the diffeomorphism symmetry of the action we expect that four of the ten equations are redundant. We expect therefore only six independent equations -- and we can use four to solve for lapse $\gamma_{rr}$ and shift $\gamma_{rA}$, and the remaining two for the graviton modes $w$ and $x$. 

Indeed  $\hat G^{rr}=0$ and $\hat G^{rA}=0$ include the lapse and shift perturbations without a radial derivative. 
We  therefore use these equations to determine the  lapse and shift components $\gamma_{rr}$ and $\gamma_{r A}$: 
\ba\label{LS1}
\gamma_{rr}\ &=&  \partial_r \left( 2\xi^r + \frac{r \Lap}{(k_t^2+k_z^2)} w\right) -w\q ,\nn\\
\gamma_{r \theta} &=&\im k_\theta \left( \xi^r + \frac{r \Lap}{2(k_t^2+k_z^2)} w \right)+ r^2 \partial_r \left( \xi^\theta + \frac{ \im k_\theta  }{2r^2(k_t^2+k_z^2)} w \right)  \, \q, \nn \\% -\im k_\theta  \lambda\frac{1}{4(k_t^2+k_z^2)}\q , \nn\\
\gamma_{r t} &=& \im k_t  \left( \xi^r + \frac{r \Lap}{2(k_t^2+k_z^2)} w \right)  + \partial_r \left( \xi^t  - \frac{\im k_t}{2(k_t^2+k_z^2)} w \right)   - \im \sqrt{\frac{2 }{\Lap^3}}\frac{k_\theta k_z}{r^2} \, x \, \q, \nn \\%- \im k_t \lambda\frac{1}{4(k_t^2+k_z^2)} \q \nn \\
\gamma_{r z} &=& \im k_z  \left( \xi^r + \frac{r \Lap}{2(k_t^2+k_z^2)} w \right) + \partial_r \left( \xi^z  - \frac{\im k_z}{2(k_t^2+k_z^2)} w \right)   + \im \sqrt{\frac{2 }{\Lap^3}}\frac{k_\theta k_t}{r^2} \, x \, \q . % - \im k_z \lambda\frac{1}{4(k_t^2+k_z^2)} \q ,
\ea
Here we have made use of the expansion 
\be\label{expansion01}
\gamma_{AB}=2r \delta^{\theta}_{A}\delta^\theta_B\xi^r + \im k_A\xi_B + \im k_B \xi_A+ wW_{AB}+xX_{AB}\,\,.
\ee

Using the solutions (\ref{LS1}) for lapse and shift in the remaining equations $\hat G^{AB}=0$, we find that these six equations reduce to two equations for the $w$- and $x$-mode respectively:  
\ba \label{gravitonsEquations}
&&\frac{d^2}{d r^2} w + \frac{1}{r} \frac{d}{d r}  w - \Lap w \,=\, 0 \q ,\nn \\
&& \frac{d^2}{d r^2} x + \frac{1}{r} \frac{d}{d r}  x - \left( \Lap + \frac{(k_t^2+k_z^2)}{r^2\Lap} \left(1 - \frac{3k_\theta^2}{r^2\Lap} \right) \right) x \,=\,0 \q .
\ea
Note that  $\Lap=r^{-2} k_\theta^2+k_t^2 +k_z^2 $ includes  an $r$-dependence.

The equation for the $w$-graviton mode are related to the modified Bessel differential equation and hence has a general solution in terms of the modified Bessel functions\footnote{${\cal I}_{k_\theta}(r)$ and ${\cal K}_{k_\theta}(r)$ are the modified Bessel function of order $k_\theta$. } 
\be\label{wsol}
w = c_1 \, {\cal I}_{|k_\theta|}\left(r\sqrt{k_t^2+k_z^2}\right)+ c_2 \, {\cal K}_{|k_\theta|}\left(r\sqrt{k_t^2+k_z^2}\right) \q ,
\ee
where $c_1$ and $c_2$ are constants.  

The equation for the $x$-mode can be related to the equation for the $w$-mode: Expressing the $x$-mode as $x = \frac{1}{\sqrt{\Lap}}  \frac{d}{dr} q$ and using this in the $x$-mode equation we find
\be
\frac{1}{\sqrt{\Lap}} \left( \frac{d}{dr} + \Lap \frac{d}{dr} \frac{1}{\Lap}  \right) \left( \frac{d^2}{d r^2}  + \frac{1}{r} \frac{d}{d r}  - \Lap \,  \right)q\,=\,0  \q,
\ee
where the differential operator in the second bracket does coincide with the one defining the equation for the $w$-mode. 

 Thus, the general solution for the $x$-mode differential equation is given by
\be\label{xsol} 
x = \frac{1}{\sqrt{\Lap}} \left( c_3 \, \frac{d}{d r} {\cal I}_{|k_\theta|}\left(r\sqrt{k_t^2+k_z^2}\right)+ c_4 \,\frac{d}{d r} {\cal K}_{|k_\theta|}\left(r\sqrt{k_t^2+k_z^2}\right) \right) \q .
\ee 
where $c_3$ and $c_4$ are constants.

\subsection{The second order Hamilton-Jacobi functional}\label{2OHJF}

Here we will compute the second order HJF, which we obtain by evaluating the action (\ref{S2O}) on solutions. As the bulk time vanishes for solutions of (\ref{eom01}), we can restrict our attention to the boundary terms appearing in (\ref{S2O}). Using the expansion (\ref{expansion01}) of the metric perturbations, and the solutions (\ref{LS1}) for the lapse and shift components $\gamma_{rr}$ and $\gamma_{rA}$ the boundary terms in (\ref{S2O}) become
%\ba\label{S2OB}
%-\kappa S^{(2)}_{\rm |_{sol}} \,=\,
%\int_{\partial {\cal M}}
% \frac{\epsilon}{4}\bigg( && \xi^r \tilde\Delta \, \xi^r   +  \, \xi^A \tilde{\cal D}_{AB}  \xi^B \,\, \,+\, \nn\\
% &&2 \left(r \Lap \xi^r + 2\im k_A\xi^A \right) w + 2 \im \sqrt{\frac{2}{\Lap}} \frac{k_\theta}{r} (k_z\xi^t - k_t\xi^z) x    \,\,+  \nn \\
%&&\frac{\Lap}{2(k_t^2+k_z^2)}(r^2 \Lap -1) w^2 -  r\,w \partial_r w - \left( \frac{1}{2} + \frac{k_\theta^2}{2r^2\Lap} \right)x^2   - \frac{r}{2} x \partial_r x\,\,\bigg)\, ,\q\q
%\ea
%where we use $\tilde\Delta = 2(k_t^2 + k_z^2)$, $ \tilde{\cal D}_{AB} = 2(k_t^2 + k_z^2)h_{AB}$. 
%{\red
 \ba\label{S2OB}
-\kappa S^{(2)} \,\underset{\text{sol}}{=} \,
 \frac{1}{2} \int_{\p \cal M} d^3y \, \epsilon \,\sqrt h 
&& \!\!\!\! \!\!  \bigg(  \xi^r \Delta \,  \xi^r   +  \,  \xi^A {\cal D}_{AB}  \xi^B \,\, \, + \, \nn\\
 &&2 \left(  2 K   \p_A \xi^A + \xi^r \Lap  \right)  w + 2  \sqrt{\frac{-2}{r^2\Lap}} K(  \p_z \xi_t  -  \p_t \xi_z) \p_\theta x    \,\,+  \nn \\
&& \frac{\Lap w}{ \Delta}(r^2 \Lap -1)  w +  \,w \, \p_r  w - \frac {x}{2 r} \left( 1 - \frac{\p_\theta^2}{r^2\Lap}  \right) x   - \frac{1}{2} x \, \p_r  x\,\,\bigg)\, ,\q\qq 
\ea 
where $\Delta = -\tfrac{2}{r} (\p_t^2 + \p_z^2), {\cal D}_{AB} = -\tfrac{2}{r} (\p_t^2 + \p_z^2) h_{AB}, \,K = \tfrac1r$ and $\Lap = -(\tfrac{1}{r^2}\p_\theta^2 +\p_t^2 + \p_z^2)$.  
In terms of Fourier modes we have for each boundary component\footnote{We remind the reader that the contribution from the mode $(k_\theta,k_t,k_z)=(0,0,0)$ is discusses in appendix \ref{zeromode}} 
\ba\label{S2OB}
-\kappa S^{(2)} \,\underset{\text{sol}}{=} \,
 \frac{\epsilon}{4} \sum_{k_\theta, k_t',k_z'}
&& \!\!\!\! \!\! \!\!\!\! \bigg( \bar \xi^r \tilde\Delta \,  \xi^r   +  \, \bar \xi^A \tilde{\cal D}_{AB}  \xi^B \,\, \,+\, \nn\\
 &&   2 \im \sqrt{\frac{2}{r^2\Lap}} \bar x ( k_z k_\theta \xi_t  - k_t k_\theta \xi_z )  -2\bar w \left(r \Lap \xi^r   + \im   k_A \xi^A \right)     \,\,+  \nn \\
&&\bar w \frac{\Lap}{\tilde \Delta}(1- r^2 \Lap )  w +   r  \, \bar w \, \p_r  w - \bar x \left( \frac{1}{2} +  \frac{k_\theta^2}{2r^2\Lap} \right) x   - \frac{r}{2} \bar x \, \p_r  x\,\,\bigg)+ cc\, ,\q\qq 
\ea
where we defined $\tilde\Delta = 2(k_t^2 + k_z^2)$, $ \tilde{\cal D}_{AB} = 2(k_t^2 + k_z^2)h_{AB}$. The bar indicates $\bar \xi(r,k_\theta, k_t',k_z) = \xi (r,-k_\theta, -k_t',-k_z')$ and  $cc$ stands for complex conjugation.

We see that the contributions to the second order HJF split into three types: (a) one part of the terms is quadratic in the perturbations describing diffeomorphisms, (b) another part mixes diffeomorphism and graviton sector, (c) the third part has terms quadratic in the graviton perturbations.
Only the part quadratic in the graviton perturbations does involve radial derivatives, and thus the solution to the graviton equation of motions.

\subsubsection{The diffeomorphism sector}

Perturbations describing diffeomorphisms do lead to contributions in the HJF, which here are given by
\ba\label{HJF2Oa} 
 -\kappa S^{(2)}_{{\rm HJa}}  &=& \frac12 \int_{\p \cal M} d^3y\, \epsilon\, \sqrt{h}\Big( \xi^r\Delta \, \xi^r  -  \, \xi^A {\cal D}_{AB}  \xi^B  \Big)  \nn\\&=& \sum_{\text{bdry comp}}    \frac{\epsilon}{4}  \sum_{k_\theta, k_t',k_z'} \,  \Big(  \bar \xi^r \tilde\Delta \,  \xi^r   +  \, \bar \xi^A \tilde{\cal D}_{AB}  \xi^B  \Big) + cc \, .\q\q %\frac{1}{2} (k_t^2+k_z^2)  \bigg[ \xi^a(r_2)\xi_a(r_2) - \xi^a(r_1)\xi_a(r_1) \bigg] \,  \q .
\ea

The part quadratic in the boundary normal diffeomorphisms can be understood to arise from the deformation of the boundary, induced by this diffeomorphisms. 
%{\color{blue} But why terms quadratic in boundary tangential diffeo's. Invariance of the full HJF under tangential diffeo's? }

This contribution to the HJF, which arises from diffeomorphism induced deformation of the background metric, can be computed for quite a wide class of backgrounds and boundaries, see Appendix \ref{AppHJF}.  For these cases one can also find a (dual) boundary field theory \cite{Hol4D,SBF}-- defined on each boundary component separately -- whose (second order) Hamilton-Jacobi functional agrees with (\ref{HJF2Oa}). This dual boundary field theory describes how the boundary (component) is embedded into the background solution \cite{Hol4D,SBF}.  This illustrates that these contributions to the HJF are localized to the boundary components.

Type (b)  contributions to the HJF mix the diffeomorphism and the graviton sector  
\ba\label{HJF2Ob}
-\kappa S^{(2)}_{\rm HJb} &=& \sum_{\text{bdry comp}}  \frac{\epsilon}{4} \sum_{k_\theta, k_t',k_z'} \,   \bigg(2 \left(r \Lap \xi^r + \im k_A\xi^A \right) \bar w + 2 \im \sqrt{\frac{2}{\Lap}} \frac{k_\theta}{r} (k_t\xi^z - k_z\xi^t) \bar x \bigg) + cc \,.\,\,
\ea 
These terms are also localized onto each  boundary component. 
%{\color{blue} Can we explain origin of these terms.}

\subsubsection{The graviton sector}

The contributions to the second order action in (\ref{S2OB}), which are quadratic in the graviton modes, involve radial derivatives.  To evaluate these terms we need to fix the constants $c_1,c_2$ and $c_3,c_4$ in the graviton solutions (\ref{wsol}) and (\ref{xsol}), respectively. To this end we have to solve the (Dirichlet) boundary value problem, so that these constants become functions of the boundary values of the graviton modes $w$ and $x$ respectively. 

We start with the case of two boundaries and abbreviate the boundary values as
\ba \label{gravitonBCs}
w(r_{\rm in}) = w_{\rm in}, \q w(r_{\rm out}) = w_{\rm out}, \q x(r_{\rm in}) = x_{\rm in}, \q x(r_2) = x_{\rm out} . \nn
\ea 
The constants $c_1,c_2$ and $c_3,c_4$ can be readily computed to be
 \ba\label{SolC}
c_1 &=&  \frac{ w_{\rm in}\, {\cal K}_{|k_\theta|}(\hat r_2) - w_{\rm out}\, {\cal K}_{|k_\theta|}(\hat r_{\rm in}) }{ {\cal I}_{|k_\theta|}(\hat r_{\rm in})\, {\cal K}_{|k_\theta|}(\hat r_2) - {\cal I}_{|k_\theta|}(\hat r_{\rm out})\, {\cal K}_{|k_\theta|}(\hat r_{\rm in}) }   \nn \\ 
c_2 &=&  \frac{ w_{\rm out}\, {\cal I}_{|k_\theta|}(\hat r_{\rm in}) - w_{\rm in}\, {\cal I}_{|k_\theta|}(\hat r_{\rm out}) }{  {\cal I}_{|k_\theta|}(\hat r_{\rm in})\, {\cal K}_{|k_\theta|}(\hat r_{\rm out}) - {\cal I}_{|k_\theta|}(\hat r_{\rm out})\, {\cal K}_{|k_\theta|}(\hat r_{\rm in})  } \nn \\
c_3 &=& \frac{ x_{\rm in} \, \sqrt{\Lap(r_{\rm in})} \frac{d}{dr_{\rm out}}{\cal K}_{|k_\theta|}(\hat r_2) - x_{\rm out} \, \sqrt{\Lap(r_{\rm out})} \frac{d}{dr_{\rm in}}{\cal K}_{|k_\theta|}(\hat r_{\rm in})  }{ \frac{d}{dr_{\rm in}}{\cal I}_{|k_\theta|}(\hat r_{\rm in})\frac{d}{dr_{\rm out}}{\cal K}_{|k_\theta|}(\hat r_{\rm out}) - \frac{d}{dr_{\rm out}}{\cal I}_{|k_\theta|}(\hat r_2)\frac{d}{dr_{\rm in}}{\cal K}_{|k_\theta|}(\hat r_{\rm in}) } \\
c_4 &=& \frac{ x_{\rm out} \, \sqrt{\Lap(r_{\rm out})} \frac{d}{dr_{\rm in}}{\cal I}_{|k_\theta|}(\hat r_{\rm in}) - x_{\rm in} \, \sqrt{\Lap(r_{\rm in})} \frac{d}{dr_{\rm out}}{\cal I}_{|k_\theta|}(\hat r_{\rm out})   }{ \frac{d}{dr_{\rm in}}{\cal I}_{|k_\theta|}(\hat r_{\rm in})\frac{d}{dr_{\rm out}}{\cal K}_{|k_\theta|}(\hat r_{\rm out}) - \frac{d}{dr_{\rm out}}{\cal I}_{|k_\theta|}(\hat r_{\rm out})\frac{d}{dr_{\rm in}}{\cal K}_{|k_\theta|}(\hat r_{\rm in})  } \nn
\ea   
where we denote $\hat r_i:= \sqrt{(k_t^2 + k_z^2)} r_i$ and $ \Lap(r_i) := \tfrac{k_\theta^2}{r_i^2} + k_t^2 + k_z^2$ for $i \in \{ {\rm in},\,{\rm out}\}$. 

Using (\ref{SolC})  for the quadratic graviton part in (\ref{S2OB}) we obtain the graviton contribution to the HJF 
 \ba\label{HJFgraviton1}
 -\kappa S^{(2)}_{\text{HJc}} (r_{\rm out},r_{\rm in}) &=&\frac{1}{4}   \sum_{k_\theta,k'_t,k'_z}  \bigg( f_{{\rm in}} \, \bar w_{{\rm in}} w_{{\rm in}} + f_{{\rm out}} \, \bar w_{{\rm out}}w_{{\rm out}} + f_{\rm io} \, \bar w_{{\rm in}} w_{\rm out} + \nn \\
 && \qq \qq  \qq \q  g_{{\rm in}} \,\bar  x_{{\rm in}}x_{{\rm in}} + g_{{\rm out}} \, \bar x_{{\rm out}} x_{{\rm out}}  + g_{\rm io} \, \bar x_{{\rm in}} x_{\rm out} \bigg) + cc \, . \qq \q 
\ea
where
\ba
f_{\rm io}(r_{\rm in},r_{\rm out},k) &=& \frac{2}{ {\cal I}_{|k_\theta|}(\hat r_{\rm in})\, {\cal K}_{|k_\theta|}(\hat r_{\rm out}) - {\cal I}_{|k_\theta|}(\hat r_{\rm out})\, {\cal K}_{|k_\theta|}(\hat r_{\rm in})  } \q \nn \\
f_{\rm in}(r_{\rm in},r_{\rm out},k) &=& r_{\rm in} \frac{\partial}{\partial r_{\rm in}} \log \left( f_{\rm io}(r_{\rm in},r_{\rm out},k) \right) + \frac{\Lap(r_{\rm in})}{2(k_t^2+k_z^2)}(r_{\rm in}^2 \Lap(r_{\rm in}) -1) \q \nn \\
f_{\rm out}(r_{\rm in},r_{\rm out},k) &=& - f_{\rm in}(r_{\rm out},r_{\rm in},k) \nn \\
g_{\rm io}(r_{\rm in},r_{\rm out},k) &=& \frac{1}{\sqrt{\Lap(r_{\rm in})\Lap(r_{\rm out})}} \frac{4}{ \frac{d}{dr_{\rm in}}{\cal I}_{|k_\theta|}(\hat r_{\rm in})\, \frac{d}{dr_{\rm out}}{\cal K}_{|k_\theta|}(\hat r_{\rm out}) - \frac{d}{dr_{\rm out}}{\cal I}_{|k_\theta|}(\hat r_{\rm out}) \, \frac{d}{dr_{\rm in}}{\cal K}_{|k_\theta|}(\hat r_{\rm in}) } \nn \\
g_{\rm in}(r_{\rm in},r_{\rm out},k) &=&  \frac{1}{2 } r_{\rm in} \frac{\partial}{\partial r_{\rm in}} \log \left( g_{\rm io}(r_{\rm in},r_{\rm out},k) \right) - \frac{1}{2} \left( 1 + \frac{k_\theta^2}{r_{\rm in}^2 \Lap(r_{\rm in})} \right) \nn \\
g_{\rm out}(r_{\rm in},r_{\rm out},k) &=& - g_{\rm in}(r_{\rm out},r_{\rm in},k)   \q .
\ea

In the case where we have only an outer boundary at $r = r_{\rm out}$, we will require that the graviton solutions are smooth at the origin, $r=0$.
By definition, the modified Bessel function $ {\cal K}_{|k_\theta|}(r)$  diverges\footnote{ The point $r=0$ is a regular singular point for the Bessel equation.} at $r=0$, thus to get smoothness, we set the constants $c_2$ and $c_4$ appearing in the graviton solutions \eqref{wsol},\eqref{xsol} to zero. We therefore, only need to determine the constants $c_1$ and $c_3$. Using the boundary conditions $w(r_{\rm out}) = w_{\rm out}, x(r_{\rm out}) = x_{\rm out}$, we obtain the  solutions:
\ba\label{GravOutSol}
w(r) = \frac{  {\cal I}_{|k_\theta|}(\hat r) }{  {\cal I}_{|k_\theta|}(\hat r_{\rm out}) } w_{\rm out}\, \, , \qq x(r) = \sqrt{ \frac {\Lap(r_{\rm out}) }{\Lap(r) } } \frac{  {\cal I}_{|k_\theta -1 |}(\hat r) +  {\cal I}_{|k_\theta +1 |}(\hat r) }{  {\cal I}_{|k_\theta -1 |}(\hat r_{\rm out}) +  {\cal I}_{|k_\theta +1 |}(\hat r_{\rm out} ) } x_{\rm out} \, \q , \qq %x_{\rm{}_b}\,
\ea
where $\hat r = r\sqrt{(k_t^2 + k_z^2)}$ and we assume $(k_t^2 + k_z^2) \neq 0$.  %These solutions are only valid for the modes $|k_\theta| \geq 2$ and $(k_t^2 + k_z^2) \neq 0$. 
 The mode $(k_\theta,k'_t,k'_z)=(0,0,0)$ requires special treatment,  see Appendix \ref{zeromode}.

With the solutions \eqref{GravOutSol} the contribution to the second order HJF of the solid torus coming from the quadratic graviton part is given by 
\ba %\sum_{ \substack {k_\theta \geq 2\\ \viol |k_t| \geq 1,|k_z|\geq 1 }} 
 -\kappa S^{(2)}_{\text{ HJ-o-c}} (r_{\rm out}) &=&\frac{1}{4}  \sum_{k_\theta,k'_t,k'_z}   \bigg(  \tilde f_{{\rm out}} \, \bar w_{{\rm out}} w_{{\rm out}}  +  \tilde g_{{\rm out}} \, \bar  x_{{\rm out}} x_{{\rm out}}   \bigg)  + cc\, ,\qq
\ea
where
\ba \label{HJoutCoeff}
\tilde f_{\rm out}( r_{\rm out},k) &=& r_{\rm out} \frac{d}{d  r_{\rm out}} \left( \log \left( {\cal I}_{|k_\theta |}(\hat r_{\rm out} )\right)  \right) + \frac{\Lap(r_{\rm out})}{\tilde \Delta}( 1 - r_{\rm out}^2 \Lap(r_{\rm out}) ) \q \nn \\
\tilde g_{\rm out}(r_{\rm out},k) &=& - \frac{1}{2 } r_{\rm out}  \frac{d}{d  r_{\rm out}} \left( \log \left(  {\cal I}_{|k_\theta -1 |}(\hat r_{\rm out} ) +  {\cal I}_{|k_\theta +1 |}(\hat r_{\rm out} )  \right) \right) - \frac{1}{2} \left( 1 + \frac{2 \, k_\theta^2}{r_{\rm out}^2 \Lap(r_{\rm out})} \right)  \, .\qq \q
\ea

\section{Hamilton-Jacobi functional via recursion relations} \label{HJRec}
 
Working with path integrals for regions with boundaries, we can `glue' the path integrals for two neighbouring regions. For our case, having the path integrals ${\cal Z}(r_2,r_1)$ and ${\cal Z}(r_3,r_2)$ for  regions with radial coordinates $r\in[r_1,r_2]$ and $r\in[r_2,r_3]$, respectively, this means that we can obtain ${\cal Z}(r_3,r_1)$ by integrating over the data associated to the shared boundary
\ba\label{rec1}
{\cal Z}(r_3,r_1) &=&\int {\cal D}\mu(\gamma_{AB}(r_2)) \,  \,  {\cal Z}(r_3,r_2) {\cal Z}(r_2,r_1) \q\  ,
\ea
 where ${\cal D}\mu(\gamma_{AB}(r))$ is a measure over the induced metric perturbations on the  $r=\text{const.}$ hypersurface. On the classical level we can `glue' the HJF's for these two regions and obtain the HJF for $[r_3,r_1]$:
 \ba\label{rec2}
 S^{(2)}_{{\rm HJ}} (r_3,r_1)&=& \underset{\gamma_{AB}(r_2)}{\text{extremize}} \left( S^{(2)}_{{\rm HJ}} (r_3,r_2)+S^{(2)}_{{\rm HJ}} (r_2,r_1) \right)  \q ,
 \ea
where we extremize the sum of the HJF's over the  induced metric perturbations $\gamma_{AB}(r_2)$. 

These convolution properties motivate to define and compute the HJF and the path integral via a discretization: the input is here the path integral for a region with radial coordinate $r\in[r_i,r_{i+1}]$ where $(r_{i+1}-r_i)$ are small. One requires smallness as one most often can only provide an approximation to the path integral, but hopes that the approximation error are small for small regions. Note that for a gravitational theory, this smallness for the radial coordinate could be seen as a rather meaningless requirement, as the physical lengths of the $[r_i,r_{i+1}]$ segments are determined by the metric, which is a variable. Thus the segments could become arbitrarily large. This turns out to be a key problem for the discretization of diffeomorphism invariant theories \cite{DittrichReview12,DittrichReview14}.

One can however use the convolution properties (\ref{rec1}) and (\ref{rec2}), and starting from an initial guess ${}^0\!{\cal Z}(r_{i+1},r_i))$  and ${}^0\!S^{(2)}_{{\rm HJ}} (r_{i+1},r_i)$, compute ${}^{1}\!{\cal Z}(r_{i+2},r_i)$ and ${}^{1}\!S^{(2)}_{{\rm HJ}} (r_{i+2},r_i)$. If we can do this computation for general values for the radii $r_j$, we can see that operation as either refining the discretization or as coarse graining. With the refining interpretation, ${}^n\!{\cal Z}(r_{\rm out},r_{\rm in}))$ and  ${}^n\!S^{(2)}_{{\rm HJ}} (r_{\rm out},r_{\rm in})$ give better and better approximation for a region with fixed radial interval $[r_{\rm in}, r_{\rm out}]$. 

In the limit $n\rightarrow \infty$ we can expect to obtain the continuum limit  for the HJF and path integral respectively. ${}^\infty\!{\cal Z}(r_{\rm out},r_{\rm in})$ and  ${}^\infty\!S^{(2)}_{{\rm HJ}} (r_{\rm out},r_{\rm in})$ are satisfying the  fixed point conditions
\ba\label{rec3}
{}^\infty\!{\cal Z}(r_{\rm out},r_{\rm in}) &=&\int {\cal D}\mu(\gamma_{AB}(r_{\rm sub})) \,  \, {}^\infty\! {\cal Z}(r_{\rm out},r_{\rm sub}) {}^\infty\!{\cal Z}(r_{\rm sub},r_{\rm in}) \q\  ,
\ea
and
\ba\label{rec4}
{}^\infty\! S^{(2)}_{{\rm HJ}} (r_3,r_1)&=& \underset{\gamma_{AB}(r_2)}{\text{extremize}} \left( {}^\infty\! S^{(2)}_{{\rm HJ}} (r_3,r_2)+{}^\infty\! S^{(2)}_{{\rm HJ}} (r_2,r_1) \right) 
\ea
respectively. One can actually attempt to directly solve the fixed point conditions and in this way obtain (a certain part of) the HJF. This strategy has been proposed in \cite{BDS11} and has been successfully tested for the harmonic and anharmonic oscillator. 

This strategy applies only to the part of  the HJF, which describe the propagating degrees of freedom, that is in our case the graviton modes. Indeed, for the second order HJF described in section \ref{2OHJF},  we only obtain non-trivial fixed point relations for part (c). The contributions (a) and (b) localize to one boundary component, and come from an outer and inner boundary with different signs, and thus cancel out for the glued boundaries.  The fixed point condition (\ref{rec4}) is therefore automatically satisfied for these contributions. 

Let us illustrate how this procedure  \cite{BDS11}  works for the construction of the second order HJF, restricted to a graviton mode $w$ with fixed momenta $(k_\theta,k_t,k_z)$. 
One starts with an ansatz for the second order HJF
\ba\label{rec5}
 -\kappa {}^0\!S^{(2)}_{{\rm HJw}} (k,r_{\rm out},r_{\rm in}) &=&\frac{1}{4}     \bigg[ {}^0\!f_{{\rm in}}(r_{\rm out},r_{\rm in}) \, w_{{\rm in}}^2 + {}^0\!f_{{\rm out}}(r_{\rm out},r_{\rm in}) \, w_{{\rm out}}^2 + {}^0\!f_{\rm io}(r_{\rm out},r_{\rm in}) \, w_{{\rm in}} w_{\rm out}  \bigg]\, ,\nn\\
\ea
where the ${}^0\!f(r_{\rm out},r_{\rm in})$ coefficients can be for instance constructed from a discretization of the Lagrangian describing the $w$-mode dynamics.  The recursion relation (\ref{rec2}) for the HJF leads then to the following recursion relations for the ${}^n\!f(r_{\rm out},r_{\rm in})$ coefficient functions:
\ba\label{rec6}
{}^{n+1}\!f_{{\rm in}}(r_{\rm out},r_{\rm in})   &=&     
  {}^{n}\!f_{{\rm in}}(r_{\rm sub},r_{\rm in}) -
  \frac{ \left({}^{n}\!f_{{\rm io}}(r_{\rm sub},r_{\rm in})\right)^2    }{4 \left(   {}^{n}\!f_{{\rm in}}(r_{\rm out},r_{\rm sub})    +   {}^{n}\!f_{{\rm out}}(r_{\rm sub},r_{\rm in})        \right)          }    
     \nn\\
{}^{n+1}\!f_{{\rm out}}(r_{\rm out},r_{\rm in})      &=& {}^{n}\!f_{{\rm out}}(r_{\rm out},r_{\rm sub}) -
  \frac{ \left({}^{n}\!f_{{\rm io}}(r_{\rm out},r_{\rm sub})\right)^2    }{4 \left(   {}^{n}\!f_{{\rm in}}(r_{\rm out},r_{\rm sub})    +   {}^{n}\!f_{{\rm out}}(r_{\rm sub},r_{\rm in})        \right)          } 
        \nn\\
{}^{n+1}\!f_{{\rm io}}(r_{\rm out},r_{\rm in}) &=&
-
  \frac{ {}^{n}\!f_{{\rm io}}(r_{\rm out},r_{\rm sub})  {}^{n}\!f_{{\rm io}}(r_{\rm sub},r_{\rm in})       }{2 \left(   {}^{n}\!f_{{\rm in}}(r_{\rm out},r_{\rm sub})    +   {}^{n}\!f_{{\rm out}}(r_{\rm sub},r_{\rm in})        \right)          } 
\ea
Here one can choose $r_{\rm sub}$ to be some point in the interval $(r_{\rm in},r_{\rm out})$, but we can expect to obtain a better convergence for a regular subdivision, e.g. $r_{\rm sub}=\tfrac{1}{2}(r_{\rm in}+r_{\rm out})$.  For $n\rightarrow \infty$ we obtain the fixed point condition -- which has to hold for any choice of $r_{\rm sub} \in (r_{\rm in},r_{\rm out})$.  Instead of applying the recursion relations to obtain (an approximation to) the coefficient functions ${}^\infty\! f$ one can also directly attempt to solve the fixed point conditions. For  examples see \cite{BDS11}.

We have already computed the coefficient functions $f$  and $g$, for the $w$- and $x$-modes respectively, in  (\ref{HJFgraviton1}). These coefficient functions do indeed satisfy the fixed point conditions derived from (\ref{rec6}).  In the next section we will make more direct use of the fixed point conditions in order to determine the one-loop correction.

%{\viol If we make an ansatz for the (quadratic part) path integral 
%\ba
%{}^0\!{\cal Z}_{\rm w}(r_{\rm out}, r_{\rm in}) = {}^0\!\mu_w(r_{\rm out}, r_{\rm in})  \, \exp \left(- \frac{1}{\hbar} S^{(2)}_{\rm HJw}(k, r_{\rm out}, r_{\rm in}) \right) \qq
%\ea
%then the convolution equation \ref{rec1} leads to the recursion relations \ref{rec6}, for the coefficients appearing in the HJF and also a recursion relation for the measure term given by
%\ba\label{recMeas}
%{}^{n+1} \!\mu_w(r_{\rm out}, r_{\rm in})  = \frac{ \sqrt{\pi \hbar} \,\, {}^{n}\! \mu_w(r_{\rm out}, r_{\rm sub}) \, {}^{n}\! \mu_w (r_{\rm sub}, r_{\rm in})   }{ 2 \sqrt{  \left(   {}^{n}\!f_{{\rm in}}(r_{\rm out},r_{\rm sub})    +   {}^{n}\!f_{{\rm out}}(r_{\rm sub},r_{\rm in})    \right)  }  } 
%\ea 
%where we have used ${\cal D}\mu( w(r_{\rm sub}) )= \d w(r_{\rm sub}) \mu(r_{\rm out},r_{\rm sub})\mu(r_{\rm sub},r_{\rm in})$ (See discussion below).
%If we consider the measure term 
%\ba
% {}^{n}\!\mu_w(r_{\rm out},r_{\rm in})  = \sqrt{\frac{-2  \, ( {}^{n}\!f_{{\rm io}}(r_{\rm out},r_{\rm in}) ) }{ \pi \hbar} }
%\ea
%then the recursion relation for the measure term \eqref{recMeas} is exactly the square root of the last term in \eqref{rec6}. 
%This implies that the fixed point solution  for the measure term can be determined from the fixed point solution of the coefficients ${}^{n}\!f_{{\rm io}}$.
%}
%~\\~\\

\section{The one-loop correction}\label{SecOLC}

Here we will utilize the convolution property (\ref{rec1})
\ba\label{rec1b}
{\cal Z}(r_3,r_1) &=&\int {\cal D}\mu(\gamma_{AB}(r_2)) \,  \,  {\cal Z}(r_3,r_2) {\cal Z}(r_2,r_1) \q\  ,
\ea
to determine the one loop correction $M(r_2,r_1)$  for the partition function  
\ba
{\cal Z}(r_2,r_1) \approx  \left(\prod_k M(k,r_2,r_1)\right)  \,\, \exp\left(     -\tfrac{1}{\hbar}\left(S^{(0)}_{\rm HJ} (r_2,r_1)+S^{(1)}_{\rm HJ} (r_2,r_1)+ S^{(2)}_{\rm HJ} (r_2,r_1) \right) \right)
\ea
 for the toroidal annulus. ${\cal D}\mu(\gamma_{AB}(r))$ is a measure over the induced metric perturbations on the $r=\text{const.}$ hypersurface.  We exclude the mode $(k_\theta,k_t,k_z)=(0,0,0)$ from the discussion, it can maximally contribute a function of the radius to the one-loop correction.  Here we absorb non-trivial measure factors into the partition function, that is into $M(k,r_2,r_1)$.   Thus  we can assume that the measure   ${\cal D}\mu(\gamma_{AB}(r))$ is just the standard Lebesgue measure for each  component of the induced metric
\ba
{\cal D}\mu(\gamma_{AB}(r)) \,=\, \prod_{A\leq B}  \prod_k  \, d\gamma_{AB}(k) \q .
\ea

Next we will parameterize the metric components $\gamma_{AB}$ via the diffeomorphism parameters $(\xi^r,\xi^A)$  and the graviton modes $(w,x)$, see (\ref{Dif1F}) and (\ref{gravi1}). The corresponding transformation of the measure is given by
\ba\label{nov6.4}
 \prod_{A\leq B} d\gamma_{AB}(k)&=& 2\sqrt{2} r^4 (k_t^2 +k_z^2) \sqrt{\Lap}\,\, d\xi^r \left(\prod_A d\xi^A \right) dw dx \q .
\ea
%Mathematica File MeasureTrafoDeterminant.nb
As we have chosen $W_{AB}$ and $X_{AB}$ in (\ref{gravi1a}) to be, modulo constants, orthonormal vectors spanning the graviton sector, we can understand the determinant to arise from the transformation acting on the diffeomorphism sector.  

We can now discuss separately the integration over the diffeomorphism modes and the integration over the graviton modes.

\subsection{Gauge sector}

As we discussed in section \ref{HJRec}, gluing the HJF's for two regions $[r_1,r_2]$ and $[r_2,r_3]$ along $r_2$, the contributions coming from the diffeomorphism sector drop out, and we are only left with a non-trivial integration over the graviton modes.

Thus the diffeomorphism parameters $(\xi^r,\xi^\theta,\xi^t,\xi^z)$ are indeed gauge parameters, which parametrize non-compact gauge orbits.  To obtain a  finite partition function we set the integration over these orbits to 1. To this end we will assume that the measure over the gauge orbits is of the form
\ba
{\cal D}_{go}(\gamma)&=& q(r)  \prod_k \left(d\xi^r \prod_A d \xi^A\right) \q ,
\ea
where we allow for a free parameter function $q(r)$.
 
 We thus have
 \ba
 {\cal D}\mu(\gamma_{AB}(r)) \,=\, {\cal D}_{go}(\gamma) \times \prod_k \frac{1}{q(r)}     2\sqrt{2} r^4 (k_t^2 +k_z^2) \sqrt{\Lap}\, \, \,dw dx   \q .
 \ea
 
 Setting the integration over the gauge orbits to 1 means that we need to consider only a reduced integral with measure
 \ba
  {\cal D}\mu_{\rm red}(\gamma_{AB}(r))\,=\, \prod_k \frac{1}{q(r)}     2\sqrt{2} r^4 (k_t^2 +k_z^2) \sqrt{\Lap}\, \, \,dw dx   \q .
 \ea
 We nevertheless have a non-trivial one-loop correction resulting from the diffeomorphism modes. These result from the factor $\frac{1}{q(r)}     2\sqrt{2} r^4 (k_t^2 +k_z^2) \sqrt{\Lap}$, which we obtained in (\ref{nov6.4})  from the transformation of $\gamma_{AB}$ to the $(\xi^r,\xi^A,w,x)$ parametrization. 
 
We therefore split the one-loop correction $M=D\times G$ into a contribution $D$ associated to the diffeomorphisms and a contribution $G$ associated to the gravitons.  Requiring the convolution property (\ref{rec1b}) to hold,  leads to the following condition on $M$
\ba
D(k,r_3,r_1) G(k,r_3,r_1)&=&    2\sqrt{2} \frac{1}{q(r_2)}r_2^4 (k_t^2 +k_z^2) \sqrt{\Lap(r_2)}  \,\,D(k,r_3,r_2) D(k,r_2,r_1) \times \nn\\ 
&&\q\q F(k,r_3,r_2,r_1) G(k,r_3,r_2)G(k,r_2,r_1)
\ea
where $F(k,r_3,r_2,r_1)$ results from the integration over the graviton modes $(w,x)$ and will be discussed further below.  Demanding that the diffeomorphism and graviton contributions separate, we obtain the condition
\ba\label{GravMRec}
G(k,r_3,r_1) = F(k,r_3,r_2,r_1) G(k,r_3,r_2)G(k,r_2,r_1)
\ea
for the graviton modes and 
\ba
D(k,r_2,r_1)\,=\, \frac{\sqrt{q(r_1)q(r_2)} }{2\sqrt{2}(k_t^2+k_z^2)\, r^2_2 r^2_1\, \Delta_b^{\frac{1}{4}}(r_2)\Delta_b^{\frac{1}{4}}(r_1)} 
\ea
for the diffeomorphism modes.

\subsection{Graviton sector}

The recursion relation (\ref{GravMRec}) describes the one-loop correction associated to the graviton sector. The factor $F$ results from a Gaussian integration, defined by the second order contribution to the HJF by the $w$- and $x$ graviton modes, see (\ref{HJFgraviton1}).  Inserting this factor we obtain the fixed point condition
\ba
G(k,r_3,r_1) = \sqrt \frac{\pi \hbar\kappa}{4(f_{{\rm in}}(r_3,r_2)  + f_{{\rm out}}(r_1,r_2) ) }  \sqrt \frac{\pi \hbar\kappa}{4(g_{{\rm in}}(r_3,r_2)  + g_{{\rm out}}(r_1,r_2) ) }  G(k,r_3,r_2)G(k,r_2,r_1) \, ,\qq
\ea
where $f_{{\rm in}},f_{{\rm out}}, g_{{\rm in}},g_{{\rm out}}$ are the coefficients appearing in the second-order HJF  \eqref{HJFgraviton1}. This condition is solved by
\ba
G(k,r_2,r_1) = \sqrt{\frac{-2  \,  f_{{\rm io}}(r_2,r_1)  }{ \pi \hbar \kappa} } \sqrt{\frac{-2  \,  g_{{\rm io}}(r_2,r_1) }{ \pi \hbar\kappa} } \q .\qq
\ea
where
\ba\label{dec5.13}
f_{\rm io}(r_{\rm in},r_{\rm out},k) &=& \frac{2}{ {\cal I}_{|k_\theta|}(\hat r_{\rm in})\, {\cal K}_{|k_\theta|}(\hat r_{\rm out}) - {\cal I}_{|k_\theta|}(\hat r_{\rm out})\, {\cal K}_{|k_\theta|}(\hat r_{\rm in})  } \q \nn \\
g_{\rm io}(r_{\rm in},r_{\rm out},k) &=& \frac{1}{\sqrt{\Lap(r_{\rm in})\Lap(r_{\rm out})}} \frac{4}{ \frac{d}{dr_{\rm in}}{\cal I}_{|k_\theta|}(\hat r_{\rm in})\, \frac{d}{dr_{\rm out}}{\cal K}_{|k_\theta|}(\hat r_{\rm out}) - \frac{d}{dr_{\rm out}}{\cal I}_{|k_\theta|}(\hat r_{\rm out}) \, \frac{d}{dr_{\rm in}}{\cal K}_{|k_\theta|}(\hat r_{\rm in}) }   \, . \nn\\
\ea

\subsection{Summary}
The one-loop correction for the toroidal annulus is therefore given by
\ba\label{Mfinal}
\prod_k M(k,r_2,r_1) \,=\, \prod_k \frac{\sqrt{q(r_1)q(r_2)}}{2\sqrt{2}(k_t^2+k_z^2)r^2_2 r^2_1\Delta_b^{\frac{1}{4}}(r_2)\Delta_b^{\frac{1}{4}}(r_1)}
\sqrt{\frac{-2  \,  f_{{\rm io}}(r_2,r_1,k)  }{ \pi \hbar\kappa} } \sqrt{\frac{-2  \,  g_{{\rm io}}(r_2,r_1,k) }{ \pi \hbar\kappa} } \q . \nn\\
\ea
Here $q(r)$ parametrizes a choice for the measure on the gauge orbits, and $f_{{\rm io}}$ and $g_{{\rm io}}$ are defined in (\ref{dec5.13}). We remind the reader that
\ba
k_t:=\frac{2\pi }{\beta_t} ( k'_t-\frac{\gamma_t}{2\pi} k_\theta), \qq  k_z:=\frac{2\pi }{\beta_z} ( k'_z-\frac{\gamma_z}{2\pi} k_\theta)
\ea
where $k_\theta,k'_z,k'_t\in \mathbb{Z}$ and $\Delta_b=k_\theta^2/r^2+k_t^2 + k_z^2$ depend on the moduli parameters $\gamma_t$ and $\gamma_z$. We have a further dependence on the moduli parameters through the arguments $\hat r=\sqrt{k_t^2+k_z^2}\,r$ which appear for the Bessel functions in (\ref{dec5.13}). For rational moduli parameters $\gamma_t/2\pi$ and $\gamma_z/2\pi$ there might exist combinations  $(k_\theta,k'_z,k'_t )\neq(0,0,0)$ for which $k_t=k_z=0$ and one  thus encounters singularities in the one-loop correction. A similar behaviour occurs for the three-dimensional version of thermal twisted flat space \cite{Hol1,HolPR}. These singularities result from a vanishing of the determinant of the Hessian of the action. But this is not caused by a gauge symmetry, rather, the linearized equations of motions do not have solutions for some subset of (linearized) boundary data \cite{Hol4D,HolPR}. The non-perturbative results on the three-dimensional case in \cite{HolPR}  indicate that these issue can be cured  by replacing the Dirichlet boundary conditions by a semi-classical boundary wave function.

\section{Discussion}\label{Disc}

Here we computed for the first time a perfect action for a system including gravitons. 
The fixed point conditions that characterize the perfect discretization lead to a new method  to determine the one-loop correction for a certain class of manifolds with boundaries. We illustrated this  for the toroidal annulus. The one-loop correction for the solid torus can be obtained by gluing to a toroidal annulus with very small inner radius a solid torus with the same very small inner radius. The partition function for the latter has been computed in \cite{Hol4D}, using a discretization for which the radial extension is just given by one radial lattice unit. This allowed to capture subtle topological effects which occur in the limit $r\rightarrow 0$, see \cite{Hol4D} for a detailed discussion.

Perfect actions restore the symmetries that have been broken by the discretization process. In the case here we do restore reparametrization invariance in the radial direction for the discretization of the background. That is gluing two toroidal annuli we can make one annulus thicker and the other thinner without changing the resulting partition function.  Likewise we can refine or coarsen the radial discretization of the annulus or solid torus, without changing the partition function. This restoration of reparametrization invariance is needed to allow for a consistent perturbative framework on the lattice, which goes beyond linear order \cite{DittrichReview12}. It does also allow for a derivation of a Hamiltonian constraint \cite{DH} for each Fourier mode, describing the dynamics in radial direction for infinitesimal radial (discretization) steps. These Hamiltonian constraints are captured by construction \cite{DH} the same dynamics as the (perfect) path integral with finite radial steps. 

 The choice of path integral measure for discrete gravity has been intensively discussed, e.g. \cite{ Meas}. 
  Constructing the one-loop correction and therefore the (one-loop) path integral measure via fixed point conditions does also avoid discretization ambiguities, and fixes otherwise free parameters in the measure \cite{Meas1,SFMeas}. 
 The resulting expression (\ref{Mfinal}) for the measure can be hardly guessed, rather it has to be computed from the dynamics of the system. 
 
 Previous work \cite{Meas1} has shown that the one-loop measure in simplicial gravity has to be necessary non-local, that is, it does not factorize over simplices. Here we encounter also non-locality as we are using a Fourier transformation in the $(\theta,t,z)$-directions. Considering a discretization of a slab of our space-time into a number of toroidal annuli, the measure does however factorize over the toroidal annuli. That is, we do have locality in radial direction, even if we do impose a momentum cut-off and thus have a discretization for all four directions.  This can be understood  from the fact that the Fourier basis modes decouple for the linearized dynamics and a momentum cut-off thus amounts to an exact truncation.  Generalizing to non-linear dynamics and keeping a cut-off for the momenta one would however expect non-local couplings for the perfect action. A way out is to use  truncations which are derived from an alternative basis, that provides a similar decoupling of the non-linear dynamics as the Fourier basis for the linear basis. Such bases can be found by using methods \cite{DittrichCC} inspired from tensor network coarse graining algorithms \cite{TNWG,TNWB}.
 
If we do not impose a Fourier mode cut-off, our result provides the computation of the one-loop partition function for a manifold with boundaries, where we also allow fluctuations for the boundary. For the latter we have seen that it is important to also consider the effects of the gauge modes on the boundary, which not only contribute to the Hamilton-Jacobi functional, but also to the one-loop correction. Here we considered a manifold with three-dimensional boundaries. It would be interesting to also consider manifolds with boundaries and corners. Here, a generalized version of the recursion relations based on gluing several manifolds (or building blocks) at once, can be employed.

\section*{Acknowledgements}

SKA is supported by an NSERC grant awarded to BD. This work is  supported  by  Perimeter  Institute  for  Theoretical  Physics.  Research
at Perimeter Institute is supported in part by the Government of Canada through the Department
of Innovation, Science and Economic Development Canada and by the Province of Ontario through
the Ministry of Colleges and Universities.

\appendix

\section{Second order expansion of the action}\label{App2ndorder}

The second order gravitational action with boundary term (\ref{S2O}) is given by \cite{Hol1,SBF}
\ba\label{Action2ndApp}
-\kappa  S^{(2)}
&=& 
\frac{1}{2} \int_{\cal M} d^d x  \sqrt{g} \, \,\gamma_{a b} \left(   
  V^{abcd} \,
  \gamma_{cd} \,\,+ \,\,
  \tfrac{1}{2}  \,
  G^{abcdef }
   \, \nabla_c \nabla_d \gamma_{ef}  \right) \,+ \nn\\
   &&
  \frac{1}{2} \int_{\p \cal M} d^{(d-1)} y   \sqrt{h}\,\epsilon \,
\gamma_{ab}  \left(\,(B_1)^{abcd}\gamma_{cd} +   \, (B_2)^{abecd} \nabla_e \gamma_{cd} \right)  
\ea
where  
\ba \label{SOTensors}
V^{abcd} &=& \frac{1}{2} \left[ \frac{1}{2}\left(R-2\Lambda \right)  \left( g^{ab}  g^{cd} - 2 g^{ac} g^{bd} \right) - R^{ab}  g^{cd}   -  g^{ab}  R^{cd} + 2\left( g^{ac} R^{bd}+ g^{bc} R^{ad}   \right) \right] \qq \nn \\ 
G^{ab cd ef} &=&  g^{ab} g^{ce}g^{df} +g^{ad} g^{bc}g^{ef} + g^{ae} g^{bf}g^{cd} - g^{ab} g^{cd}g^{ef} - g^{ad} g^{bf}g^{ce} - g^{af} g^{bd}g^{ce}    \nn \\
B_1^{abcd} &=&  \frac{1}{2} ( K h^{ab}-K^{ab} ) g^{cd} - h^{ac} h^{bd}K - h^{ab}K^{cd} + h^{ac} K^{bd}+ h^{bd}K^{ac}     \nn \\ 
B_2^{ab e cd} &=&  \frac{1}{2} \left( \left(  h^{ae} h^{bd} - h^{ab} h^{ed} \right)n^c + \left(  h^{ac} h^{be} - h^{ab} h^{ce} \right) n^d -  \left(  h^{ac} h^{bd} - h^{ab} h^{cd} \right)n^e \right). \q 
\ea

For background vacuum solutions, we can write 
\ba
V^{abcd} = \frac{\Lambda}{d-2}  \left( 2 g^{ac} g^{bd} - g^{ab}  g^{cd}  \right) \q .
\ea
The variation of the action (\ref{Action2ndApp}) with respect to the metric perturbations $\gamma_{ab}$ leads to the equations of motion
 \ba\label{EOMlambdaApp}
\hat G^{ab}:=\left(   
  V^{abcd} \,
  \gamma_{cd} \,\,+ \,\,
  \tfrac{1}{2}  \,
  G^{abcdef }
   \, \nabla_c \nabla_d \gamma_{ef}  \right)&=&0  \q .
\ea
One can show explicitly that the diffeomorphism induced perturbations $\gamma_{ab}=\nabla_a\xi_b+\nabla_b \xi_a$ satisfy (\ref{EOMlambdaApp}).  

\section{Hamilton-Jacobi functional for the diffeomorphism sector}\label{AppHJF}

In this section we will determine the second order of the Hamilton-Jacobi functional, restricted to solutions that are diffeomorphism equivalent to the  (homogeneous) background solution.  This includes the case of twisted thermal flat space, which we are considering in the main text and which will appear as somewhat special. 

Here we will consider space-times of general dimensions $d\geq 3$. The case of $d=3$ has been extensively discussed in \cite{SBF}. The higher dimensional case adds a few extra features but the discussion is quite similar to the three--dimensional case. We will therefore be very short and refer to more details and proofs to \cite{SBF,SethThesis}.

We choose the background solution to be expressed in Gaussian normal coordinates %topology $\Sigma \times \mathbb{R}$ and express the
\be\label{bgmetric}
ds^2 = g_{ab} dx^a dx^b = dr^2 + h_{AB} dy^A dy^B
\ee
where the space--time coordinates  have the form $x^a = (r, y^A)$  with a  radial coordinate $r \in \mathbb R ^+$. We assume the manifold to has one or two boundary components (in this case referred to as ``inner'' and ``outer'') at constant radius. 

We will consider background vacuum solutions that are Einstein manifolds, that is the background Riemann tensor satisfies
\ba\label{ConCurv1}
R_{abce} = \frac{4\Lambda}{(d-1)(d-2)} g_{a[c}  g_{e]b} \q .
\ea
For the background intrinsic curvature of the  $r= \rm const$ hypersurfaces, and thus the boundaries we assume that the Riemann tensor  ${} \b R^B{}_{CDE}$ is covariantly constant  $D_A \,\,  \b R^B{}_{CDE} =0 $, where with $D_A$ we denote the covariant derivative associated to the  $r= \rm const$ hypersurface. We also  assume  for  the (background) extrinsic curvature  $K_{AB}=\tfrac{1}{2}\partial_r h_{AB}$ that $\tilde \pi^{AB}:=(K^{AB}-Kh^{AB})$ is everywhere non-vanishing  on the boundary.

Here we will consider only  the diffeomorphism sector, that is perturbative solutions $\gamma_{ab}\equiv \zeta_{ab}$, which arise from an infinitesimal diffeomorphism generating vector field $\xi^a$
\ba
\zeta_{ab} = \left[{\cal L}_\xi g\right]_{ab} \q .
\ea
Due to the Gaussian form of the background metric we have for the $y^A$-components of the metric perturbations
\ba\label{AppB001}
\zeta_{AB} \,=\, \left[{\cal L}_\xi g\right]_{AB}\,=\, 2 K_{AB} \xi^r + D_A \xi_B + D_B \xi_A  \q .
\ea

In the following we will parameterize the boundary metric perturbations $\zeta_{AB}$, which we assume to be induced by an infinitesimal diffeomorphism, by the diffeomorphism generating vector field $\xi^a$ itself. In particular we will express the Hamilton--Jacobi functional as a functional of $\xi^a$. To this end it is helpful to determine $\xi^a$ as a functional of the boundary metric perturbations (projected to the diffeomorphism sector) $\zeta_{AB}$, that is to invert (\ref{AppB001}). This inversion can be obtained through
%
% Let us require that the relationship between the metric fluctuations projected onto the diffeomorphism sector $\zeta_{AB}$ and the diffeomorphism inducing vector field $(\xi^\perp, \xi^A)$ in  \eqref{Dif1} is invertible. Using the shorthand $\tilde \pi^{AB} = \left(K^{AB} - K h^{AB} \right) $ which is the de-densitized conjugate momentum $\tilde \pi^{AB} = h^{-1/2} \pi^{AB}$, we will now state our first result. 
%
%
\ba \label{xiofg}
\Pi^{AB} \zeta_{AB} &=& \Delta \, \xi^r \qq ,  \nn \\%+ 2D^B(\b R_{AB}) \xi^A \q , \nn\\
2\tilde\pi^{BC}\delta'\b\Gamma^A_{BC} &=& {\cal D}^A{}_B \xi^B + Q^{AB} D_{B}\xi^r  \qq \q \q %&& - D_B(\b R^{AB})\xi^\perp  \blue (==\pi^{BC} D^A K_{BC} )\q
\ea
where we defined
\ba\label{DefOp1}
\Pi^{AB} &=& \left( h^{AC} h^{BD} - h^{AB}h^{CD} \right) D_C D_D - \b R^{AB} \q , \nn\\
\delta'\, \b \Gamma^A_{BC} &=& \frac{1}{2} h^{AD} \left( D_B \zeta_{CD} + D_C \zeta_{BD} - D_D \zeta_{BC}  \right) \q, \nn \\
\Delta &=& 2 \tilde \pi^{AB} D_A D_B - 2\, \b R^{AB} K_{AB} \q ,\nn \\ 
{\cal D}^A{}_B &=& 2\tilde \pi^{CD}\left( D_CD_D h^A_B +  \b R^A{}_{CBD} \right) \q , \nn \\
%&=& 2\tpi^{CD} D_CD_D h^A_B + 2\left( R^{AB}K_{BD} -R K^A_D + \frac{2(d-3)\Lambda}{(d-2)}\left( K^A_D - \frac{1}{(d-1)} K h^{A}_D \right) \right)\xi^D \q   \nn\\
Q^{AB} &=& 2 \left( 2\tilde \pi^{BC}  K^A_C - \tilde \pi^{CD} K_{CD} h^{AB} \right)\,=\, 2\left(\left( \b R h^{AB} - 2\, \b R^{AB} \right) - 2\Lambda \frac{(d-3)}{(d-1)} h^{AB} \right)\, .
  %2\left(\left( \b R h^{AB} - 2 \b R^{AB} \right) - 2\Lambda \frac{(d-3)}{(d-1)} h^{AB} \right) \q .
\ea
The last equation follows from the Gauss-Codazzi relations and our assumption (\ref{ConCurv1}). In dimensions $d=3$, the background tensor $Q^{AB}$ vanishes automatically since the Ricci scalar for two dimensional surfaces  satisfies ${}^2\!R_{AB} = \tfrac{1}{2} {}^2\!R h_{AB} $.  In dimensions $d=4$ it can be non-vanishing, it is however zero for the example of thermal flat space, considered in the main text.

The second order of the Hamilton-Jacobi functional (restricted to the diffeomorphism sector) is then given by
\ba
{}^{\rm D}\!S^{(2)}_{\rm HJ}&=& \! -\frac{1}{ 2\kappa} \int_{\p \cal M} d^{(d-1)} y \; \sqrt{h}  \epsilon \,  \bigg( \xi^r (\Delta + Q^{AB}K_{AB}) \xi^r  \,+ \xi^r Q^{AB}D_A \xi_B -  \nn \qq \\ 
&& \hspace{7.5cm} \xi_A Q^{AB} D_B\xi^r -  \, \xi^A {\cal D}_{AB}  \xi^B  \bigg) \q .  \qq \q
\ea
Note that this (restricted) Hamilton-Jacobi functional is local as a functional of the diffeomorphism generating vector field components $\xi^a$ (which are however non-local functionals of the boundary induced metric perturbations $\gamma_{AB}$). In the case that $Q^{AB}=0$, one can utilize this local form to construct  a local boundary field theory of a scalar and vector field defined on the boundary, whose Hamilton-Jacobi functional does reproduce the second order gravitational Hamilton-Jacobi functional (restricted to the diffeomorphism sector).  This allows for a holographic interpretation of the diffeomorphism sector of general relativity \cite{Hol1,Hol4D,SBF, SethThesis} and as first discussed in \cite{Hol4D}, applies also to the example of four-dimensional thermal flat space discussed in the main text.

\section{The $k_\theta=k'_t=k'_z=0$ mode}\label{zeromode}

Here we will shortly discuss the special case of the  Fourier mode $k_\theta=k_t'=k_z'=0$. The equations of motion $\hat G^{ab}=0$ reduce for this case to
 \begin{align}
 \hat G^{rr} &= -\frac{1}{2r} \p_r (\gamma_{tt}+\gamma_{zz}) , \q \,\, \hat G^{rA} =0, \qq \qq \qq \qq \hat G^{\theta \theta} = -\frac{1}{2r^2}  \p_r^2 (\gamma_{tt}+\gamma_{zz})  \nn \\
\hat G^{\theta t} &= \frac{1}{2r} \p_r \left( \frac 1 r \p_r \gamma_{\theta t }\right) , \qq \q  \hat G^{\theta z} = \frac{1}{2r} \p_r \left( \frac 1 r \p_r\gamma_{\theta z }\right),  \qq \,\,\,  \hat G^{tz} = \frac{1}{2r} \p_r \left( r \p_r \gamma_{ tz }\right),  \q  \\
 \hat G^{tt} &= \frac{1}{2r} \p_r \left( \gamma_{rr} +\frac{1}{r^2} \gamma_{\theta \theta} - \frac1 r \p_r \gamma_{\theta \theta} - r \p_r \gamma_{zz} \right), \q \hat G^{zz} = \frac{1}{2r} \p_r \left( \gamma_{rr} +\frac{1}{r^2} \gamma_{\theta \theta} - \frac1 r \p_r \gamma_{\theta \theta} - r \p_r \gamma_{tt} \right)\, , \nn
 \end{align}
 where the index $A$ appearing in the first line stands for $A=\theta,t,z$.
 
 Different from the general case the shift components $\gamma_{rA}$ do not appear in the equation of motion and the components $\hat G^{rA}$ vanish automatically.  We still have one constraint equation $\hat G^{rr}=0$ -- characterized by the fact of being only a first order differential equation in the radius. It does only involve the spatial components (and neither lapse or shift), and constitutes a constraint on the boundary values of the spatial metric.

 The solutions to these equations of motion can be parametrized as follows 
 \begin{align} \label{k0sol} 
 \gamma_{r r} &=  c_0 +2 \p_r \xi^r , \q && \gamma_{r\theta}=r^2\p_r\xi^\theta, \q && \gamma_{rB}=\p_r\xi^B,  \nn \\
 \gamma_{\theta\theta}&=2r \xi^r ,\q &&
 \gamma_{tt} =  -c_1 \log(r) +c_2 , \q && \gamma_{zz} =  c_1 \log(r) + c_3, \nn\\
 \gamma_{\theta t} &= c_4 \, r^2  + c_5 ,\q && \gamma_{\theta z} = c_6 \, r^2  + c_7, \q && \gamma_{tz} = c_8 \log(r) + c_9 \qq
 \end{align}
 where $B=z,t$. 
 Here $\xi^a$ are the components of the diffeomorphism generating vector field and 
  $c_i, i\in \{ 0,\cdots,9\}$ are additional parameters appearing in the solutions.
 
Substituting the solutions \eqref{k0sol} into the second-order action we find the second-order HJF (with inner and outer boundary) given by 
  \ba
 \kappa S^{(2)}_{{\rm HJ|}_{k_\theta=0}} =&& \frac 12   \left( C_1 \, (r_{\rm out}^2-r_{\rm in}^2) + C_2 (\log(r_{\rm out})-\log(r_{\rm in})) + C_3( \log(r_{\rm out})^2- \log(r_{\rm in})^2 )\right) 
\ea
 where the $C_j$ are given by
 \ba
% C_1 &=& c_3 (3 c_1-2 c_2+c_3)+4 c_4 c_5+4 c_6 c_7+2 c_9 (c_8+2 c_9) , \q
 C_1 = 2 (c_4{}^2+c_6{}^2),\q 
 C_2 &=& c_1(c_1-2c_2+2c_3)+c_8 (c_8+4 c_9) , \q C_3 = 2 (c_1{}^2+c_8{}^2) \q . \qq
 \ea
% 
%  \ba 
% -\kappa S^{(2)}_{\rm HJ}  &=&  \int_{\p M} \frac{1}{4} \epsilon  \left(4 \left(c_4{}^2+c_6{}^2\right) r^2+c_3 (3 c_1-2 c_2+c_3)+4c_4 c_5+4 c_6 c_7+2 c_9 (c_8+2 c_9)\right) \nn \\ 
%   && \q \q -\epsilon  \log (r) \left(2 \left(c_2{}^2+c_8{}^2\right) \log (r)+c_2{}^2+2 c_3 c_2+c_8 (c_8+4 c_9)\right) 
% \ea
% Or 
 
With an inner boundary at $r=r_{\rm in}$ and an outer boundary at $r=r_{\rm out}$, Dirichlet boundary conditions determine the constants $c_i$ to be 
 \ba
 c_1 &=&-\frac{  \gamma_{tt}(r_{\rm out}) -  \gamma_{tt}(r_{\rm in})   }{  \log(r_{\rm out}) - \log(r_{\rm in}) }  \,=\, \frac{  \gamma_{z z}(r_{\rm out}) -  \gamma_{z z}(r_{\rm in})   }{  \log(r_{\rm out}) - \log(r_{\rm in}) } \q , \qq  \nn\\
 c_2&=&   \frac{   \gamma_{tt}(r_{\rm out})\log(r_{\rm in}) - \gamma_{tt}(r_{\rm in}) \log(r_{\rm out})   }{  \log(r_{\rm in}) - \log(r_{\rm out}) } 
 \,\,,\q\q
 c_3 = \frac{   \gamma_{z z}(r_{\rm out})\log(r_{\rm in}) - \gamma_{z z}(r_{\rm in}) \log(r_{\rm out})   }{  \log(r_{\rm in}) - \log(r_{\rm out}) } \q
 \ea
 where there are two independent equations determining $c_1$. This constitutes a constraint on the boundary values which results from the constraint $\hat G^{rr}=0$ above. 
We have furthermore
 \ba
 c_4 &=& \frac{   \gamma_{\theta t}(r_{\rm out}) - \gamma_{\theta t}(r_{\rm in})   }{  r_{\rm out}^2 - r_{\rm in}^2  } 
 \q , \qq c_5 =  \frac{   \gamma_{\theta t}(r_{\rm out}) r_{\rm in}^2 - \gamma_{\theta t}(r_{\rm in}) r_{\rm out}^2  }{  r_{\rm in}^2 - r_{\rm out}^2  } \q , \nn \\
  c_6 &=& \frac{   \gamma_{\theta z}(r_{\rm out}) - \gamma_{\theta z}(r_{\rm in})   }{  r_{\rm out}^2 - r_{\rm in}^2  } \q , \qq c_7 =  \frac{   \gamma_{\theta z}(r_{\rm out}) r_{\rm in}^2 - \gamma_{\theta z}(r_{\rm in}) r_{\rm out}^2  }{  r_{\rm in}^2 - r_{\rm out}^2  } \q, \nn \\
  c_8 &=& \frac{   \gamma_{t z}(r_{\rm out}) - \gamma_{t z}(r_{\rm in})   }{  \log(r_{\rm out}) - \log(r_{\rm in}) } \q , \qq c_9 = \frac{   \gamma_{t z}(r_{\rm out})\log(r_{\rm in}) - \gamma_{t z}(r_{\rm in}) \log(r_{\rm out})   }{  \log(r_{\rm in}) - \log(r_{\rm out}) }  \q. \qq
 \ea

If we have only an outer boundary the manifold will include the $r=0$ axis. Demanding that the metric perturbations are smooth around $r=0$ in cartesian coordinates, and transforming back to cylindrical coordinates \cite{SBF}, we find that for the solutions (\ref{k0sol}) some parameters have to vanish:
\ba
c_1=0 \, , \,\, c_5=0 \, ,\,\, c_7=0 \, ,\,\, c_8=0 \q .
\ea
The contribution to the Hamilton-Jacobi Functional is then given by 
\ba
 \kappa S^{(2)}_{{\rm HJ|}_{k_\theta=0}} &=& \frac{1}{4} \left(   
 (c_2-c_3)^2+c_0(c_2+c_3) + 4 c_9^2+ 4(c_4^2+c_6^2)r_{\rm out}^2       \right). \q  ,
\ea
where
\ba
c_2=\gamma_{tt}(r_{\rm out}) \, ,\,\, c_3=\gamma_{zz}(r_{\rm out}) \, ,\,\, c_9=\gamma_{tz}(r_{\rm out}) \, ,\,\,
c_4=\frac{\gamma_{\theta t}(r_{\rm out})}{r_{\rm out}^2}  \, ,\,\, 
c_6=\frac{\gamma_{\theta z}(r_{\rm out})}{r_{\rm out}^2} 
\ea
and $c_0$ can be determined as
\ba
c_0= \gamma_{rr}(r_{\rm out}) - \left[ \partial_r \frac{\gamma_{\theta\theta}(r)}{r}\right]_{r=r_{\rm out}} \q .
\ea

\providecommand{\href}[2]{#2}
\begingroup
%\raggedright
\endgroup

\end{document}